\pdfoutput=1
\documentclass[11pt,twoside,a4paper,cmspaper,final,collab]{cms-tdr}
\def\svnVersion{ef7ed65}\def\svnDate{2024/04/15}\def\cmsCernNoTag{CERN-EP-2023-249}\def\cmsCernDate{\today}\def\cmsMessage{Published in Physics Letters B as \href{http://dx.doi.org/10.1016/j.physletb.2024.138633}{\doi{10.1016/j.physletb.2024.138633}.}}
\begin{document}\cmsNoteHeader{BPH-21-005}

\newcommand{\mumu}{\ensuremath{\PGmp\PGmm}\xspace}
\newcommand{\mumupi}{\ensuremath{\PGmp\PGmm\PGpp}\xspace}
\newcommand{\mmp}{\ensuremath{\PGm\PGm\PGp}\xspace}
\newcommand{\taunu}{\ensuremath{\PGtp\PGnGt}\xspace}
\newcommand{\munut}{\ensuremath{\PGmp\PGnGt}\xspace}
\newcommand{\munu}{\ensuremath{\PGmp\PGnGm}\xspace}
\newcommand{\tautrimu}{\ensuremath{\PGt \!\to\! 3\PGm}\xspace}
\newcommand{\Wtaunu}{\ensuremath{\PW \!\to\! \PGt\PGnGt}\xspace}
\newcommand{\Wmunu}{\ensuremath{\PW \!\to\! \PGm\PGnGm}\xspace}
\newcommand{\ppWX}{\ensuremath{\Pp\Pp \!\to\! \PW + \PX}\xspace}
\newcommand{\dstaunu}{\ensuremath{\PsDp \!\to\! \taunu}\xspace}
\newcommand{\dsphipi}{\ensuremath{\PsDp \!\to\! \PGf\PGpp \!\to\! \mumupi}\xspace}
\newcommand{\BtauX}{\ensuremath{\PB \!\to\! \PGt + \PX}\xspace}
\newcommand{\BDsX}{\ensuremath{\PB \!\to\! \PsDp + \PX}\xspace}
\DeclareRobustCommand{\PBsz}{\HepParticle{\PB}{s}{0}\xspace}
\ifthenelse{\boolean{cms@external}}{\providecommand{\figwidewidth}{0.5\textwidth}}{\providecommand{\figwidewidth}{0.6\textwidth}} 
\ifthenelse{\boolean{cms@external}}{\providecommand{\figsixwidth}{0.41\textwidth}}{\providecommand{\figsixwidth}{0.45\textwidth}} 

\cmsNoteHeader{BPH-21-005}
\title{Search for the lepton flavor violating \texorpdfstring{\tautrimu}{tau to 3mu} decay in proton-proton collisions at \texorpdfstring{$\sqrt{s} = 13\TeV$}{sqrt(s) = 13 TeV}}

\author{The CMS Collaboration}

\date{\today}

\abstract{A search for the lepton flavor violating \tautrimu decay is performed using proton-proton collision events at a center-of-mass energy of 13\TeV collected by the CMS experiment at the LHC in 2017--2018, corresponding to an integrated luminosity of 97.7\fbinv. Tau leptons produced in both heavy-flavor hadron and \PW boson decays are exploited in the analysis. No evidence for the decay is observed. The results of this search are combined with an earlier null result based on data collected in 2016 to obtain a total integrated luminosity of 131\fbinv. The observed (expected) upper limits on the branching fraction $\mathcal{B}(\tautrimu)$ at confidence levels of 90 and 95\% are $2.9 \times 10^{-8}$ ($2.4 \times 10^{-8}$) and $3.6 \times 10^{-8}$ ($3.0 \times 10^{-8}$), respectively.}

\hypersetup{%
pdfauthor={CMS Collaboration},%
pdftitle={Search for the lepton flavor violating tau to 3 mu decay in proton-proton collisions at sqrt(s) = 13 TeV},%
pdfsubject={CMS},%
pdfkeywords={CMS, BSM, CLFV, charged lepton flavor violation}}

\maketitle 

\section{Introduction}
\label{sec:intro}

There are no known symmetries that would strictly forbid lepton flavor violating decays,
such as \tautrimu.
However, the only source of lepton flavor violation in the standard model (SM) of particle physics is from neutrino transitions in loops, which
implies a vanishingly small branching fraction for \tautrimu,
around $10^{-55}$~\cite{Pham:1998fq,Hernandez-Tome:2018fbq,Blackstone:2019njl}.
Various extensions of the SM predict \tautrimu decays with branching fractions
as high as $10^{-10}$--$10^{-8}$~\cite{Marciano:2008zz,Raidal:2008jk,Arganda:2005ji},
which can be probed in present and near-future experiments.

The most stringent experimental upper limit on the branching fraction of the \tautrimu decay, set by the Belle
experiment, is $\mathcal{B}(\tautrimu) < 2.1 \times 10^{-8}$ at 90\% confidence level
(\CL)~\cite{Hayasaka:2010np}.
A similar upper limit of $3.3 \times 10^{-8}$ at 90\% \CL~\cite{Lees:2010ez}
is reported by the BaBar experiment.
While Belle and BaBar operated at asymmetric electron-positron B factories,
proton-proton ($\Pp\Pp$) collisions offer another prolific source of tau leptons.
The upper limits at 90\% \CL reported by the CERN LHC experiments are $4.6 \times 10^{-8}$ from LHCb~\cite{Aaij:2014azz},
$38 \times 10^{-8}$ from ATLAS~\cite{Aad:2016wce}, and $8.0 \times 10^{-8}$ from CMS~\cite{CMS:2020kwy}.

In this Letter,
we report a search for the \tautrimu decay with the CMS experiment using data collected in 2017 and 2018,
which correspond to integrated luminosities of 38.0\fbinv and 59.7\fbinv, respectively.
The result is combined with a previously published analysis using data collected by the CMS experiment in 2016,
which corresponds to an integrated luminosity of 33.2\fbinv~\cite{CMS:2020kwy}.
Both searches exploit tau leptons produced in heavy-flavor (charm and bottom) hadron decays and
\PW boson decays.
Heavy-flavor decays produce the vast majority of tau leptons at the LHC, and the \tautrimu signal is
characterized by the presence of low transverse momentum (\pt) muons in the final state, typically
below 10\GeV, with high intrinsic background.
While the tau lepton production rate from \PW boson decays is a few orders of magnitude smaller, the
muons from the \tautrimu signal have larger \pt, are typically isolated from hadronic activity,
and are accompanied by significant missing transverse momentum in the event, resulting in a low
intrinsic background.
The background is dominated by events in which a pion or kaon misidentified as a muon is combined with two muons from a \PQb hadron
decay (usually one muon from the \PQb hadron decay and one from the subsequent \PQc hadron decay).
Tabulated results are provided in the HEPData record for this analysis~\cite{hepdata}.

\section{The CMS detector}
\label{sec:CMS}

The CMS apparatus~\cite{Chatrchyan:2008zzk} is a multipurpose, nearly hermetic detector, designed to trigger on~\cite{CMS:2020cmk,Khachatryan:2016bia} and identify electrons, muons, photons, and (charged and neutral) hadrons~\cite{CMS:2020uim,Sirunyan:2018fpa,CMS:2014pgm}. A global ``particle-flow" algorithm~\cite{Sirunyan:2017ulk} aims to reconstruct all individual particles in an event, combining information provided by the all-silicon tracker and by the crystal electromagnetic and brass-scintillator hadron calorimeters, operating inside a 3.8\unit{T} superconducting solenoid, with data from the gas-ionization muon detectors embedded in the flux-return yoke outside the solenoid.
The reconstructed particles are used to build jets and measure the missing transverse momentum (\ptvecmiss)~\cite{Sirunyan:2018pgf,CMS:2016lmd,CMS:2019ctu},
which is computed as the negative vector sum of the \pt of all the reconstructed particles in an event, with magnitude denoted as \ptmiss. 

Muons are measured in the pseudorapidity range $\abs{\eta} < 2.4$, with detection planes made using three technologies: drift tubes, cathode strip chambers, and resistive-plate chambers.
Matching measurements in the muon detectors to tracks measured in the tracker results in a relative \pt resolution, for muons with \pt up to 100\GeV, of 1\% in the barrel and 3\% in the endcaps~\cite{Sirunyan:2018fpa}.

Events of interest are selected using a two-tiered trigger system. The first level (L1), composed of custom hardware processors, uses information from the calorimeters and muon detectors to select events at a rate of around 100\unit{kHz} within a fixed latency of about 4\mus. The second level, known as the high-level trigger (HLT), consists of a farm of processors running a version of the full event reconstruction software optimized for fast processing, and reduces the event rate to around 1\unit{kHz} before data storage~\cite{Khachatryan:2016bia}. 

\section{Data and simulated samples}
\label{sec:dataset}

The data used in this analysis were collected in 2017--2018 from $\Pp\Pp$ collisions at a center-of-mass
energy of 13\TeV, and correspond to an integrated luminosity of 97.7\fbinv.
The heavy-flavor analysis uses \dstaunu and $\PBp / \PBz\!\to\!\PGt +\PX$ as signal processes,
accounting for about 60 and 40\% of the $\PGt$ lepton production from heavy-flavor decays, respectively,
and $\PsDp \!\to\! \PGf(1020)\PGpp \!\to\! \mumupi$ as a normalization process
($\PGf(1020)$ is written as \PGf in the remainder of the Letter). 
Charge-conjugated processes are implied throughout this Letter.
Simulated event samples are generated with Monte Carlo (MC) methods, using 
\PYTHIA8.226~\cite{PYTHIA} with the CP5 tune~\cite{CMS:2019csb} at leading order (LO).
The heavy-flavor decays are modeled with \EVTGEN1.6.0~\cite{EVTGEN}, and a phase space model
is used for the \tautrimu decay.
The $\PWp\to\taunu$ process is also simulated using \PYTHIA with the CP5 tune.
In addition, a smaller $\PWp\to\taunu$ sample is simulated using \MGvATNLO~\cite{Alwall:2014hca}
at next-to-LO (NLO).
All generated events are passed to a detailed \GEANTfour~\cite{GEANT} simulation of the CMS detector.
The subsequent trigger selections and event reconstruction are performed with the same
algorithms as those used with the data.
The simulated events include multiple $\Pp\Pp$ interactions within the same or nearby bunch crossings, known as pileup, and are reweighted to match the pileup distribution in data.

\section{Search for the \texorpdfstring{\tautrimu}{tau to 3mu} decay in heavy-flavor hadron events}
\label{sec:HFsearch}

The search for \tautrimu in heavy-flavor hadron decays uses low-$\pt$
dimuon or trimuon triggers.
The 2017 data were collected with L1 trigger requirements of either two oppositely charged muons
having a pseudorapidity of $\abs{\eta}<1.5$ and a separation of $\Delta R = \sqrt{\smash[b]{(\Delta\eta)^2 + (\Delta\phi)^2}} < 1.4$,
where $\Delta\eta$ and $\Delta\phi$ are the differences in pseudorapidity and azimuthal angle, respectively,
or three muons,
with the $\pt$-(sub)leading muon having $\pt > 5(3)\GeV$
(without any explicit \pt requirement for the third muon).
At the HLT, two oppositely charged muons with $\pt>3\GeV$
and an additional track with $\pt>1.2\GeV$ were required.
The three particles were fitted to a common vertex, which was required to have a vertex fit $\chi^2$ per degree of
freedom less than 8 and be separated from the beamline in the transverse plane by at least twice the uncertainty in the
distance.
The invariant mass of the three particles, assuming the muon mass for each, was required to be in
the range of 1.60--2.02\GeV, which is wide enough to record both \tautrimu and \dsphipi candidate events.
The two-dimensional pointing angle ($\alpha_{\mathrm{2D}}$),
defined as the angle in the transverse plane between the three-particle momentum vector and the vector from
the beamline to the three-particle common vertex, was required to satisfy $\cos{\alpha_{\mathrm{2D}}}>0.9$.
For the 2018 data-taking period, an L1 trigger algorithm requiring two oppositely charged $\pt>4\GeV$
muons with $\Delta R<1.2$ was added to complement the previous ones.
The HLT algorithm was changed to require three muons, but with the same \pt and vertex requirements as those in 2017.
The previous ``2 muons and 1 track'' trigger was prescaled by a factor of 20, \ie, storing only
5\% of the events selected by the algorithm,
in order to keep recording \dsphipi candidate events.

In the offline selection, three global muons~\cite{Sirunyan:2018fpa} are required. A global muon combines a muon candidate reconstructed
in the muon detectors with a track reconstructed in the silicon tracker.
The three selected muons must have $\abs{\eta}<2.4$ and the minimum \pt is 3\GeV for the leading two
muons and 2\GeV for the third, when ordered by decreasing \pt.
Any muon with $\abs{\eta}<1.2$ must have $\pt>3.5\GeV$.
The muons must be matched to the objects that triggered the event by $\Delta R<0.03$ and have a total charge
of $\pm$1.
In about 1\% of the cases, where more than one muon triplet candidate is found, the one with
the smallest trimuon vertex fit $\chi^2$ is selected.
The tracks of the three muon candidates are refitted with a constraint that they originate from a common point~\cite{kinFit}.
This improves the mass resolution by about 10\%. 
To match the trigger requirement, the two-dimensional distance of the trimuon vertex to the beamline is required to be at least twice its uncertainty.
The primary vertex (PV) is chosen
as the one with the smallest three-dimensional pointing angle ($\alpha_{\mathrm{3D}}$) between the trimuon momentum
and the vector connecting the PV and the trimuon vertex.

\begin{figure}[hbtp]
  \centering
    \includegraphics[width=\figwidewidth]{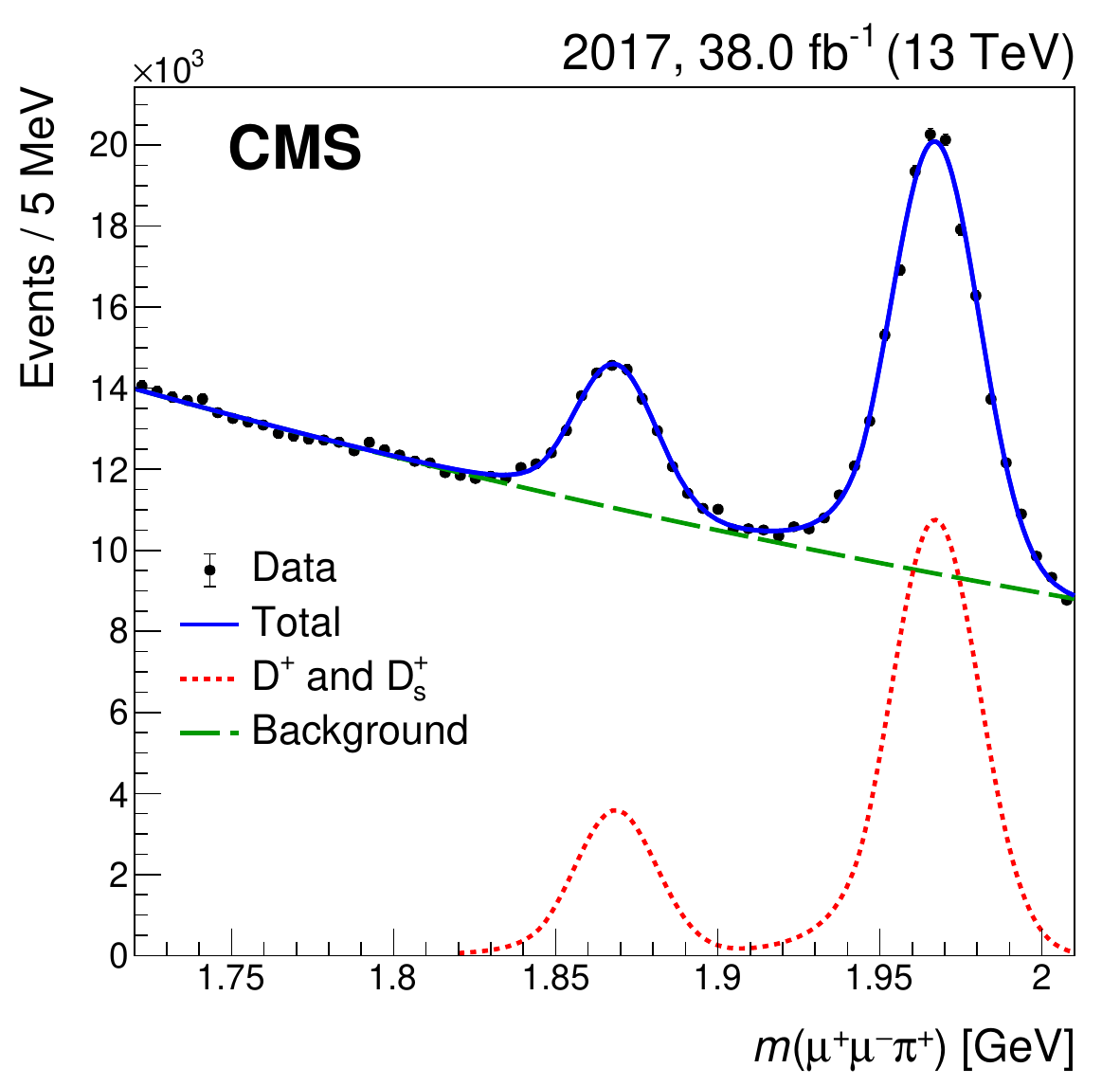}
 \caption{
The \mumupi invariant mass distribution with the fits to the sum of the \PDp (1.870\GeV) and \PsDp (1.968\GeV)~\cite{PDG2022} resonances
and the background in 2017 data.
}
\label{fig:dsphipi}
\end{figure}

The signal normalization strategy, detailed in Ref.~\cite{CMS:2020kwy}, is summarized here.
The dependence on the knowledge of D and B meson production cross sections, or trigger and selection
efficiencies, is minimized by normalizing the signal yield to the \dsphipi yield in data, separately
for 2017 and 2018.
The normalization channel \dsphipi uses the same selection criteria with the following exceptions.
Only two muons are required and they must be oppositely charged with an invariant mass between 1 and
1.04\GeV. The track associated with the pion must have $\pt>2\GeV$ and form a vertex with the two
muons with a $\chi^2$ per degree of freedom less than 5. 
Figure~\ref{fig:dsphipi} shows the \mumupi invariant mass distribution using 2017 data with
fits to the \PDp and \PsDp signal peaks using Crystal Ball functions~\cite{CrystalBallRef} for the
peaks and an exponential function for the background. 

The expected number of \tautrimu signal events from \PsDp meson decays that pass the dimuon triggers, denoted as $N_{3\PGm(\PD)}$, is related to $\mathcal{B}(\tautrimu)$ by
\ifthenelse{\boolean{cms@external}}
{
\begin{linenomath}
\begin{equation*}
\begin{aligned}
N_{3\PGm(\PD)} \, &= \, N_{\mmp}
\frac{\mathcal{A}_{3\PGm(\PD)}}{\mathcal{A}_{\mmp}} 
\frac{\epsilon_{3\PGm(\PD)}^{\text{reco}}}{\epsilon_{\mmp}^{\text{reco}}}
\frac{\epsilon_{3\PGm(\PD)}^{2\PGm\text{trig}}}{\epsilon_{\mmp}^{2\PGm\text{trig}}} \\
& \times \,
\frac{\mathcal{B}(\dstaunu)}{\mathcal{B}(\dsphipi)} \,
\mathcal{B}(\tautrimu),
\end{aligned}
\end{equation*}
\end{linenomath}
}
{
\begin{linenomath}
\begin{equation*}
N_{3\PGm(\PD)} \, = \, N_{\mmp}
\frac{\mathcal{B}(\dstaunu)}{\mathcal{B}(\dsphipi)} 
\frac{\mathcal{A}_{3\PGm(\PD)}}{\mathcal{A}_{\mmp}}
\frac{\epsilon_{3\PGm(\PD)}^{\text{reco}}}{\epsilon_{\mmp}^{\text{reco}}}
\frac{\epsilon_{3\PGm(\PD)}^{2\PGm\text{trig}}}{\epsilon_{\mmp}^{2\PGm\text{trig}}}\,
\mathcal{B}(\tautrimu),
\end{equation*}
\end{linenomath}
}
where $N_{\mmp}$ is the measured \dsphipi yield, $\mathcal{A}$ is the detector acceptance,
$\epsilon^\text{reco}$ is the selection efficiency, $\epsilon^\text{trig}$ is the trigger efficiency,
and the subscripts ${3\PGm(\PD)}$ and \mmp indicate \tautrimu and \dsphipi decays, respectively.
Similarly, the expected number of \tautrimu signal events from decays of the form \BtauX coming from the dimuon triggers, denoted as $N_{3\PGm(\PB)}$,
is related to $\mathcal{B}(\tautrimu)$ by
\ifthenelse{\boolean{cms@external}}
{
\begin{linenomath}
\begin{equation*}
\begin{aligned}
N_{3\PGm(\PB)} \, &= \,N_{\mmp}\, f\, 
\frac{\mathcal{A}_{3\PGm(\PB)}}{\mathcal{A}_{\mmp}}
\frac{\epsilon_{3\PGm(\PB)}^{\text{reco}}}{\epsilon_{\mmp}^{\text{reco}}}
\frac{\epsilon_{3\PGm(\PB)}^{2\PGm\text{trig}}}{\epsilon_{\mmp}^{2\PGm\text{trig}}}\,
\mathcal{B}(\tautrimu)\\
& \times \,
\frac{\mathcal{B}(\BtauX)}{\mathcal{B}(\BDsX) \mathcal{B}(\dsphipi)},
\end{aligned}
\end{equation*}
\end{linenomath}
}
{
\begin{linenomath}
\begin{equation*}
N_{3\PGm(\PB)} \, = \, N_{\mmp}\, f\, 
\frac{\mathcal{B}(\BtauX)}{\mathcal{B}(\BDsX) \mathcal{B}(\dsphipi)} 
\frac{\mathcal{A}_{3\PGm(\PB)}}{\mathcal{A}_{\mmp}}
\frac{\epsilon_{3\PGm(\PB)}^{\text{reco}}}{\epsilon_{\mmp}^{\text{reco}}}
\frac{\epsilon_{3\PGm(\PB)}^{2\PGm\text{trig}}}{\epsilon_{\mmp}^{2\PGm\text{trig}}}\,
\mathcal{B}(\tautrimu),
\end{equation*}
\end{linenomath}
}

where $N_{\mmp}$ is the measured \dsphipi yield, $f$ is the fraction of
$\PsDp$ mesons from \PQb hadron decays.
The fraction $f$ can be calculated as the ratio of cross sections ($\sigma$): 
$f = \sigma(\Pp\Pp\!\to\!\PB + \PX)\mathcal{B}(\BDsX)/\sigma(\Pp\Pp\!\to\!\PsDp+\PX)$.  Since \PsDp mesons produced from
\PQb hadron decays tend to decay farther from the PV than directly produced \PsDp mesons, we
validate the simulation-predicted value of $f$ by fitting the \dsphipi proper decay length, $Lm(\mumupi)/p$, distribution in data,
where $L$ is the distance between the PV and the \mumupi vertex, $m(\mumupi)$ is the \mumupi invariant mass, and $p$ is
the \mumupi momentum.
The small contributions to signal events from $\PDp\!\to\!\PGt+\PX$ and $\PBsz \!\to\! \PGt+\PX$ decays
are added by scaling the \dstaunu and
\BtauX predictions by 1.04 and 1.12, respectively, as determined from simulation. 
Systematic uncertainties equal to these corrections are assigned.

The data collected using the trimuon trigger cannot be directly normalized to \dsphipi.
The simulation predicts that the fraction of signal events triggered exclusively by the trimuon trigger is about 28\% of the events
passing the dimuon triggers. When measured from events in a $\PsDp \!\to\! \PGf\munu \!\to\! \mumu\munu$
control region, this ratio is found to be 31\%. The dimuon-triggered signal yields are therefore scaled up by 28\%, and a systematic
uncertainty of 3\% is assigned on the total yields.

Selected events are categorized based on the calculated trimuon mass resolution,
which primarily depends on the pseudorapidities of the reconstructed muon candidates.
Three mass resolution categories are introduced: $\sigma_m/m < 0.7\%$, $0.7 < \sigma_m/m < 1.05\%$, and $\sigma_m/m > 1.05\%$, where $m$ is the trimuon invariant mass and $\sigma_m$ is its uncertainty.
They are labeled A, B, and C, with average $\sigma_m$ of 12, 19, and 25\MeV, respectively.
The signal region (SR) for each category includes candidates with trimuon invariant masses within twice the average mass resolution.
The sideband region includes candidates outside the SR but with trimuon invariant masses of 1.62--2.00\GeV.
To remove $\PGf\!\to\! \PGmp\PGmm$ contributions, candidates in which two oppositely charged muons have an invariant
mass near the \PGf resonance are rejected, with the exact requirement depending on the resolution category.

\begin{figure*}[hbtp]
  \centering
    \includegraphics[width=0.48\textwidth]{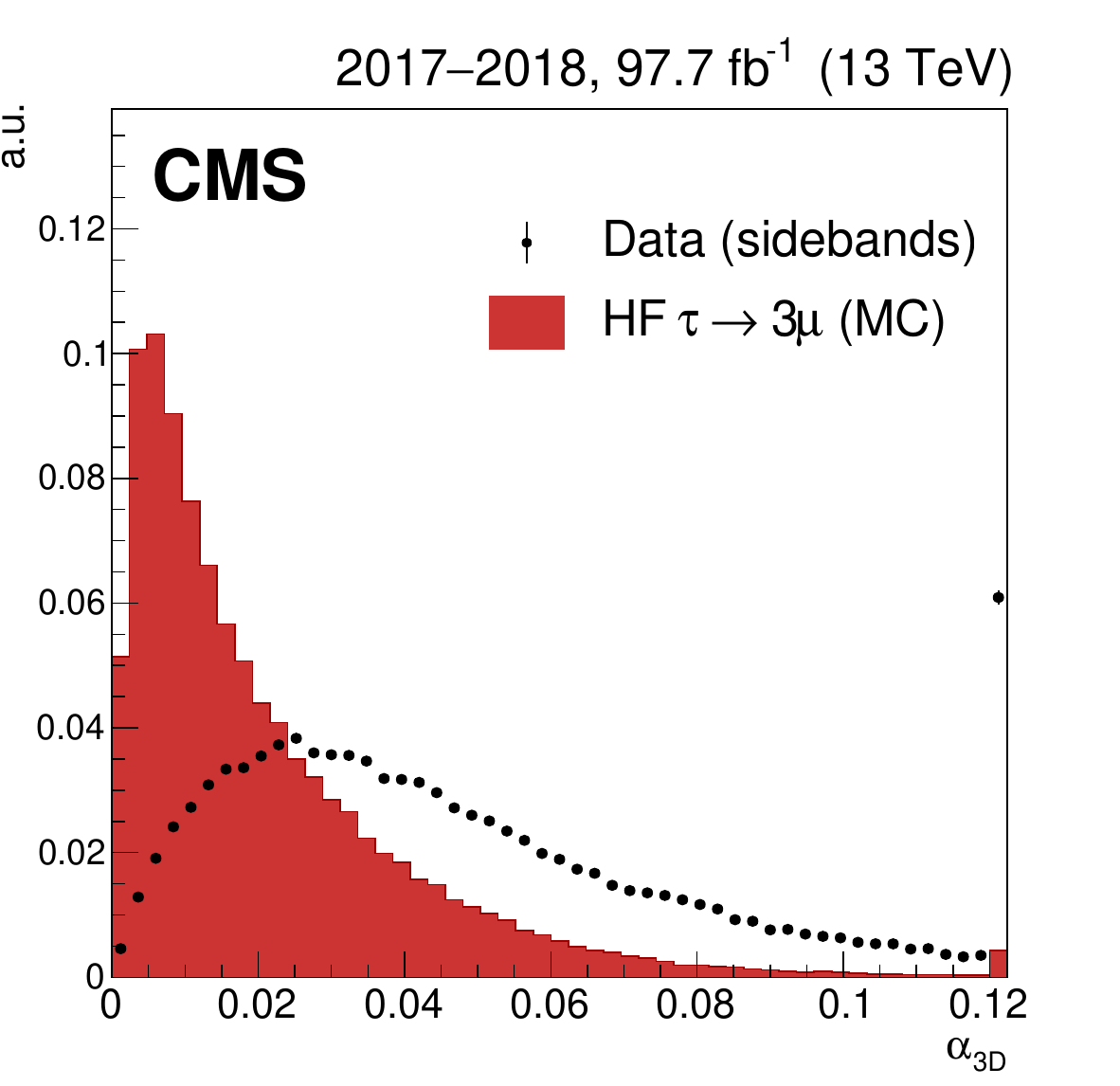}
    \includegraphics[width=0.48\textwidth]{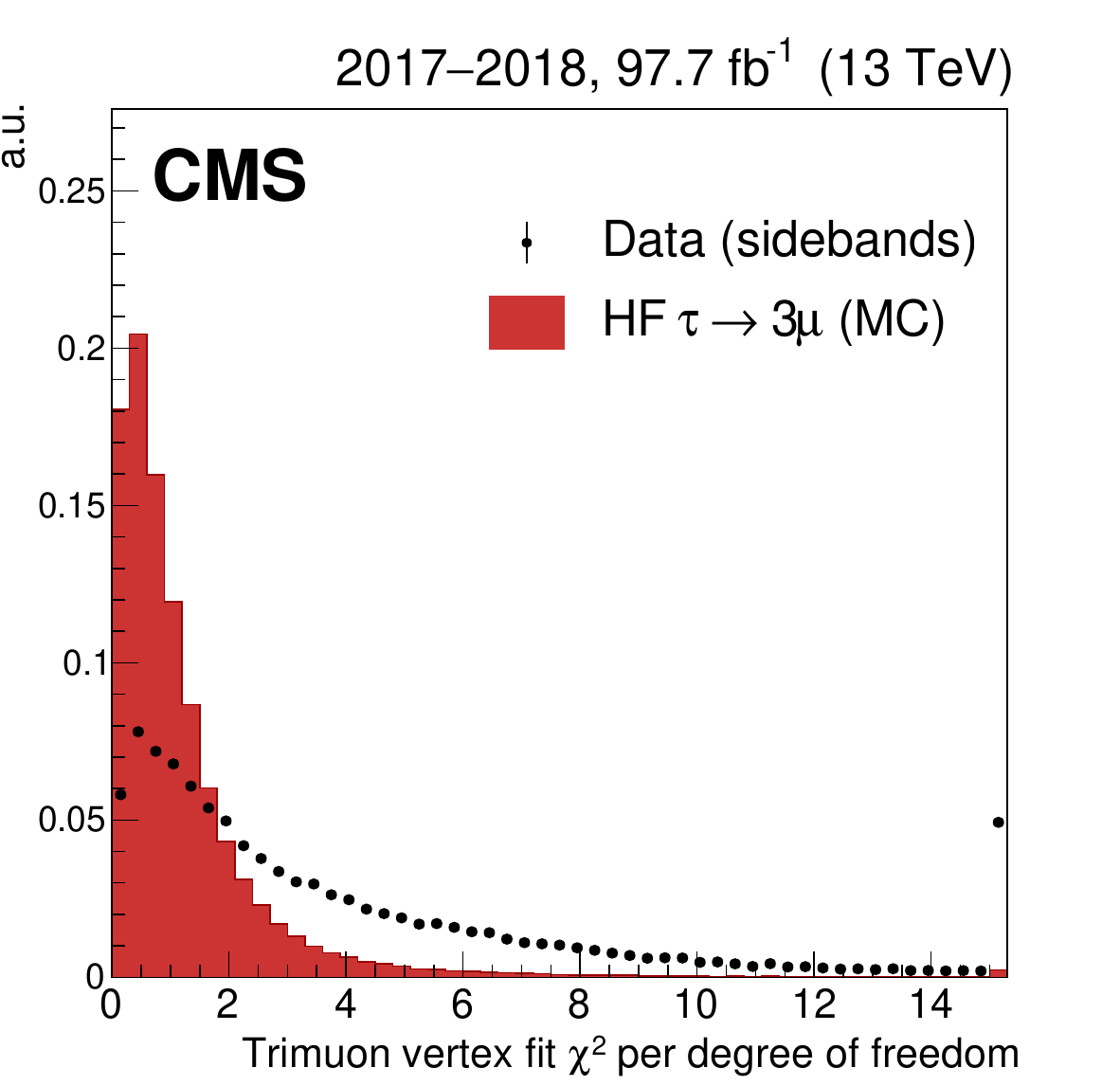}
    \includegraphics[width=0.48\textwidth]{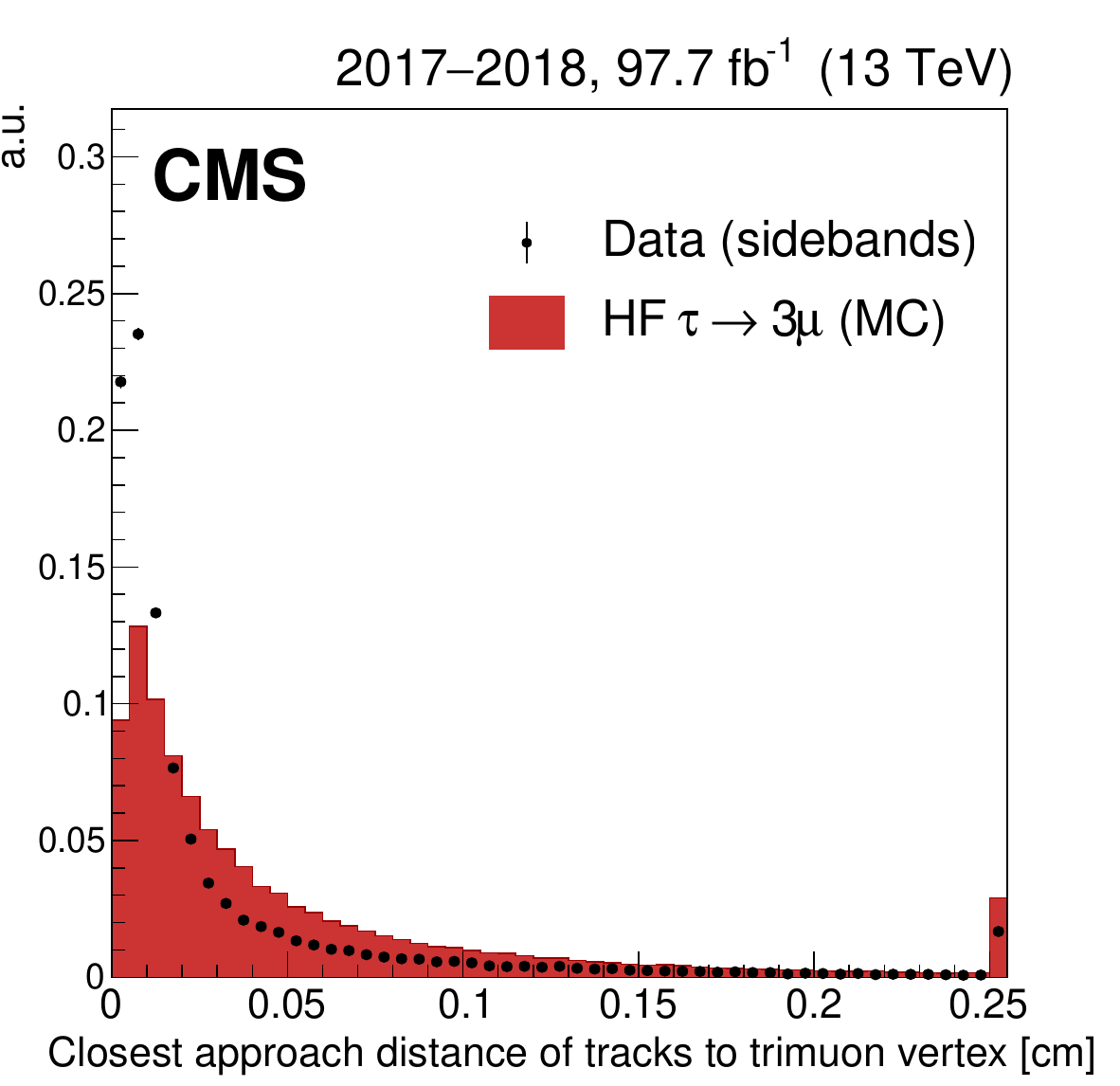}
    \includegraphics[width=0.48\textwidth]{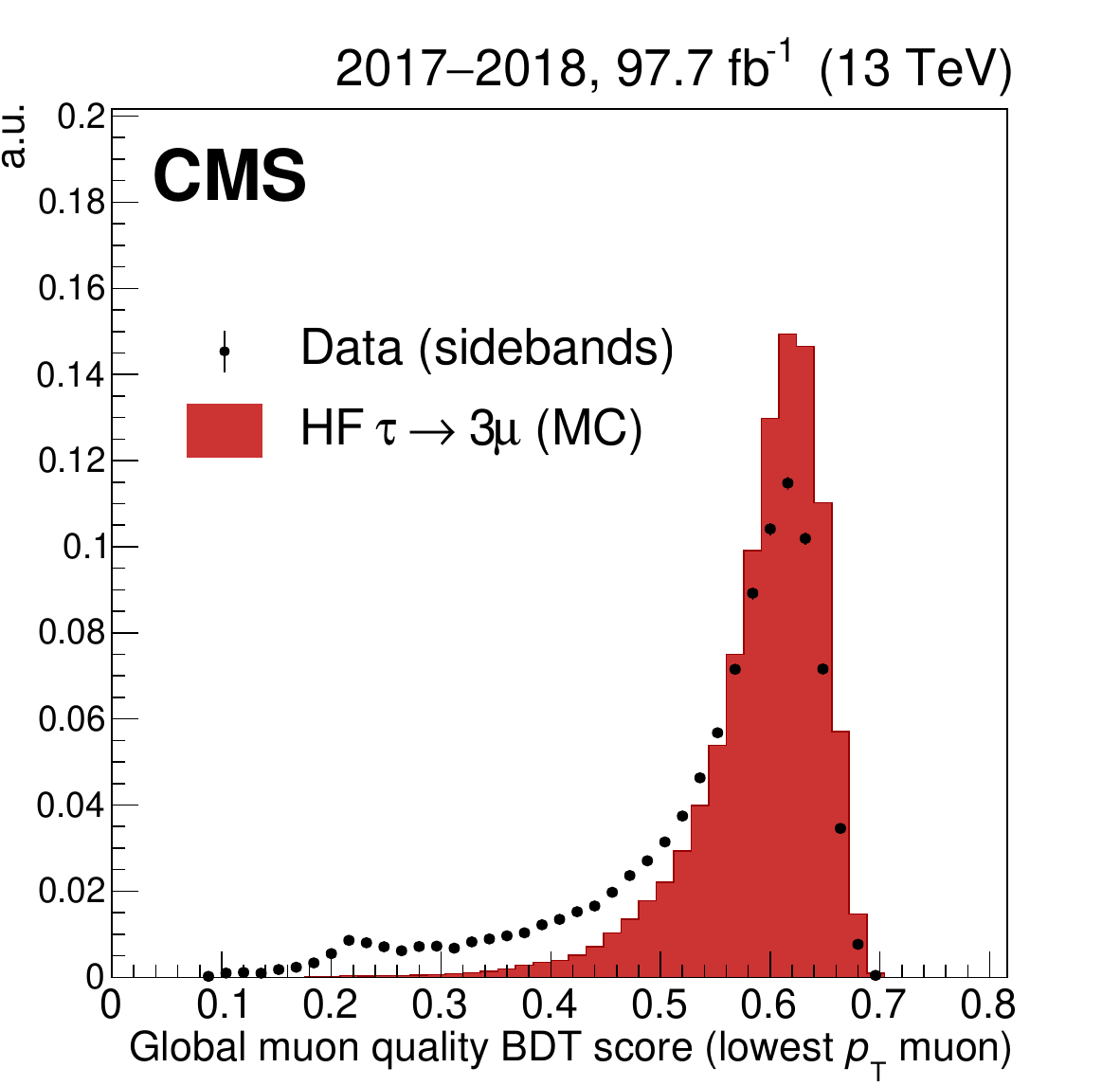}
 \caption{Signal and background distributions for the four observables with the highest discrimination
power used for the heavy-flavor (HF) analysis BDT training: $\alpha_{\mathrm{3D}}$ (upper left),
$\chi^2$ per degree of freedom of the trimuon vertex fit (upper right),
the smallest distance of closest approach to the trimuon vertex of
all the other tracks in the event with $\pt>1\GeV$ (lower left), and 
the muon reconstruction quality BDT score of the lowest \pt muon of the triplet (lower right).
The signal and background distributions are obtained respectively from MC simulation
and from the mass-sideband regions in data.
All distributions are normalized to unit area.
The rightmost bins include overflow.
}
\label{fig:hf_BDT}
\end{figure*}

A study of simulated minimum-bias events reveals that the few events that pass our selection criteria usually have
at least one muon candidate that is associated with a generated hadron.
This can occur from the decay of a pion or kaon as well as from random matching between hadron tracks in the tracker
and stubs in the muon detectors.
Therefore, a boosted decision tree (BDT) is trained~\cite{Chen:2016:XST:2939672.2939785} based on muon reconstruction quality,
using the lowest-\pt muon among the three in simulated \tautrimu events as signal,
and simulated B meson to charged pion or kaon decays as background, where hadrons wrongly identified as global muons are selected.
The muon reconstruction quality BDT makes use of observables from the silicon
tracker and muon detectors, as well as their compatibility.
After that, another BDT, referred to as the analysis BDT, is trained for each mass resolution category to improve the signal-to-background ratio, using
simulated signal events (with properly mixed D and B meson decays)
and background events from the data sideband region.
The analysis BDT training utilizes the following observables:
trimuon \pt, 
the muon reconstruction quality BDT scores of each of the three muon candidates,
additional muon quality criteria (each using the least signal-like value among the three muon candidates),
number of hits in muon detectors for each of the three muon candidates,
the normalized $\chi^2$ of the trimuon vertex fit, $\alpha_{\mathrm{3D}}$,
the distance between the PV and the trimuon vertex and its significance (defined as the distance divided by its uncertainty),
the largest and smallest values of the transverse impact parameter of the three muons with respect to the
PV, and their significances,
and two isolation observables.
The isolation observables measure the activity in terms of other particles in the vicinity of the trimuon vertex or the three muons.
The first isolation observable is the smallest distance of closest approach to the trimuon vertex of all
other tracks in the event with $\pt>1\GeV$.
The second isolation observable, which is maximized over the three muons, is defined by summing the \pt of all tracks
with $\pt>1\GeV$, and with $\Delta R < 0.3$ and a distance of closest approach below 1\mm with respect to the muon
(the tracks associated with the other two muons are excluded), and by dividing this sum
by the muon candidate \pt.
The distributions of the observables providing the largest discrimination power between signal and background 
are shown in Fig.~\ref{fig:hf_BDT}.

Each mass resolution category is divided into four subcategories based on the analysis BDT score.
While the lowest analysis BDT score
subcategory is discarded, the other three subcategories in each mass resolution category 
are retained for the statistical analysis, and assigned number labels from 1 to 3, 
such that label 1 corresponds to the highest (the most signal-enriched) BDT score subcategory.
In this way, nine event categories are defined for each data-taking year.

The boundaries of the analysis BDT subcategories are optimized to give the largest expected combined signal
significance, which is done before unblinding the signal region.
For consistency with the \PW boson analysis, described in Section~\ref{sec:Wsearch},
$\PGoP{782}\!\to\! 2\PGm$ contributions are vetoed
by requiring the invariant masses of oppositely charged muon pairs to be away from the \PGoP{782} peak by more than 10\MeV.
This requirement has no impact on the result.

\begin{figure*}[hbtp]
  \centering
    \includegraphics[width=\figsixwidth]{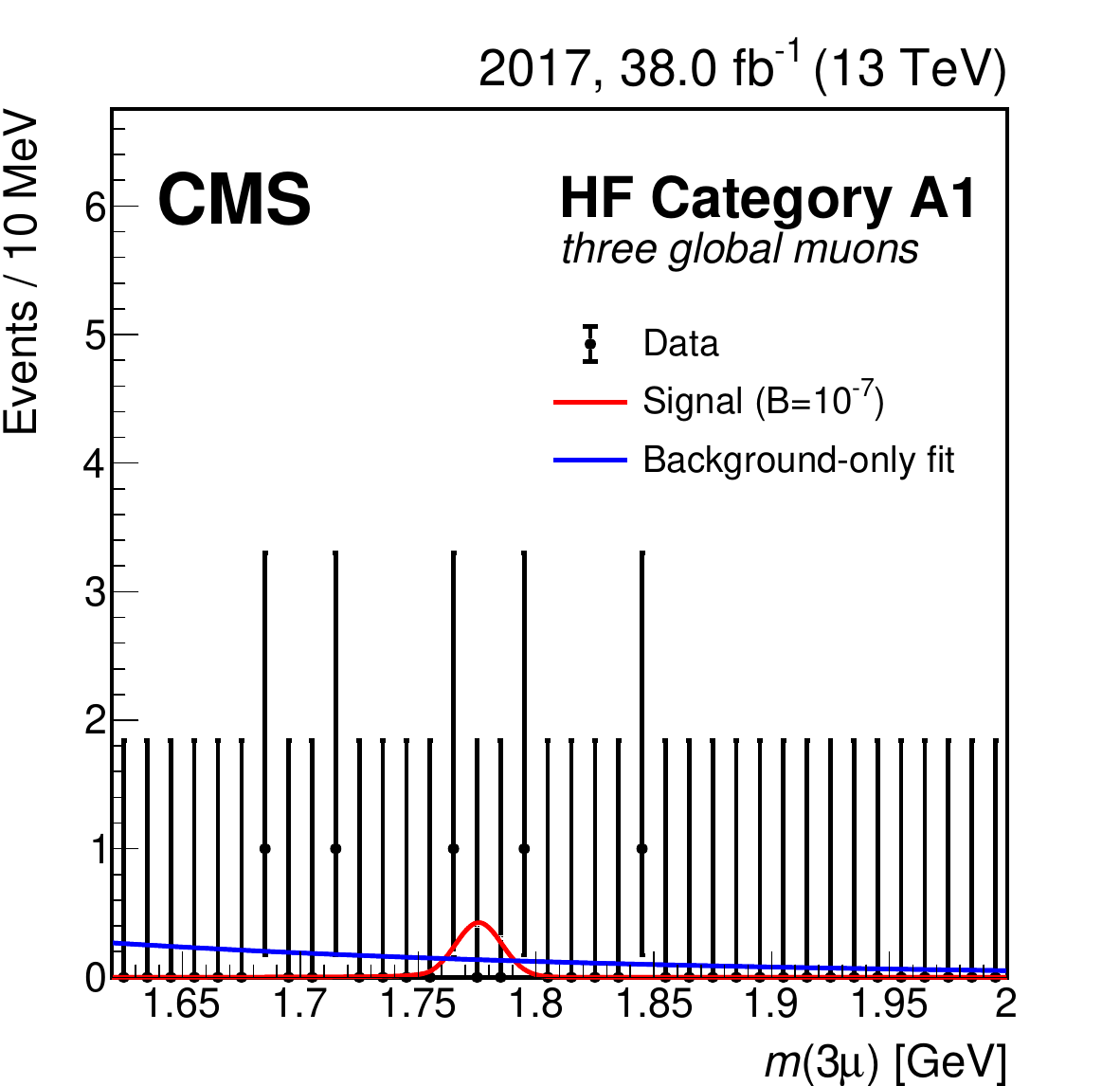}
    \includegraphics[width=\figsixwidth]{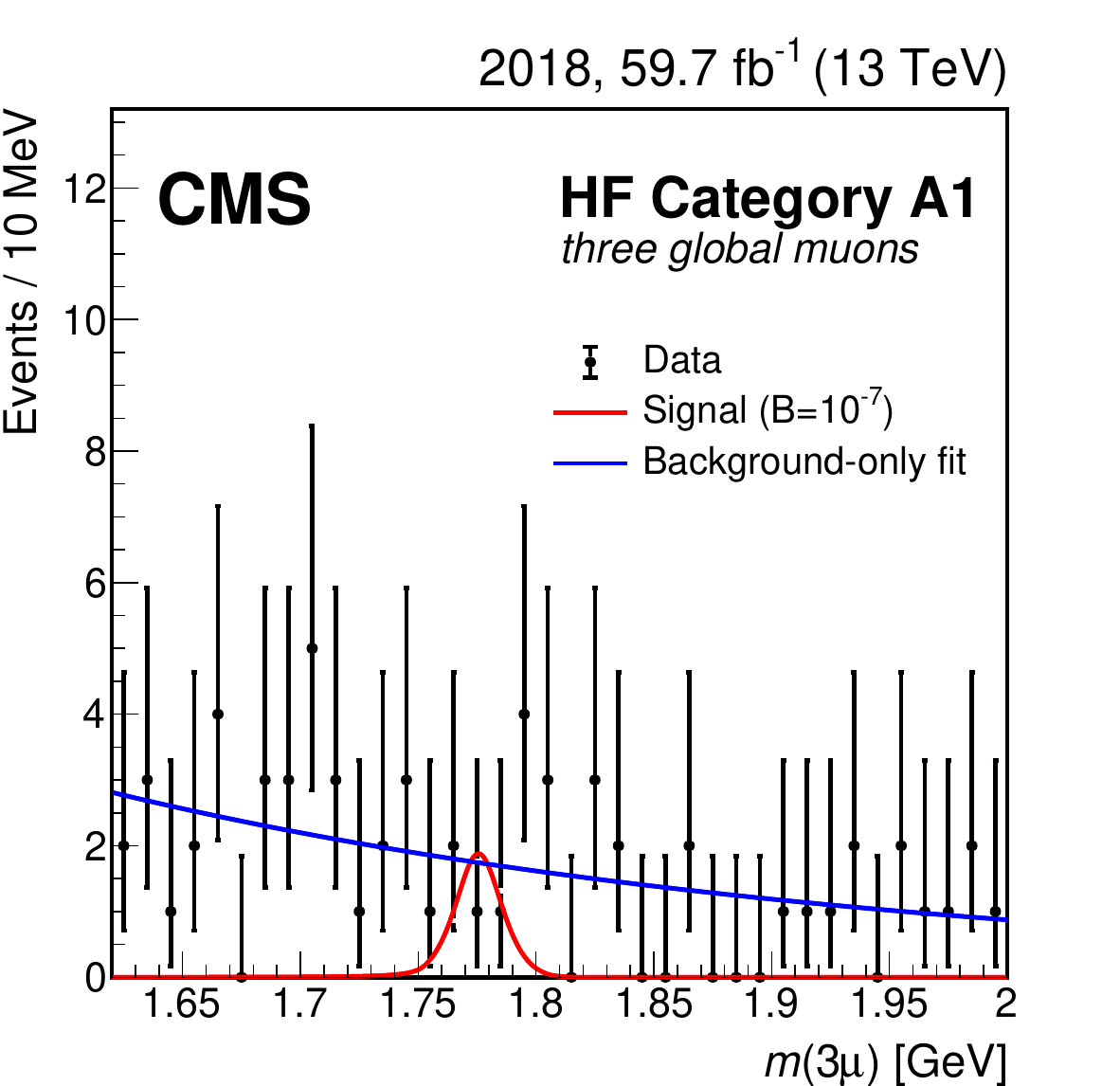}
    \includegraphics[width=\figsixwidth]{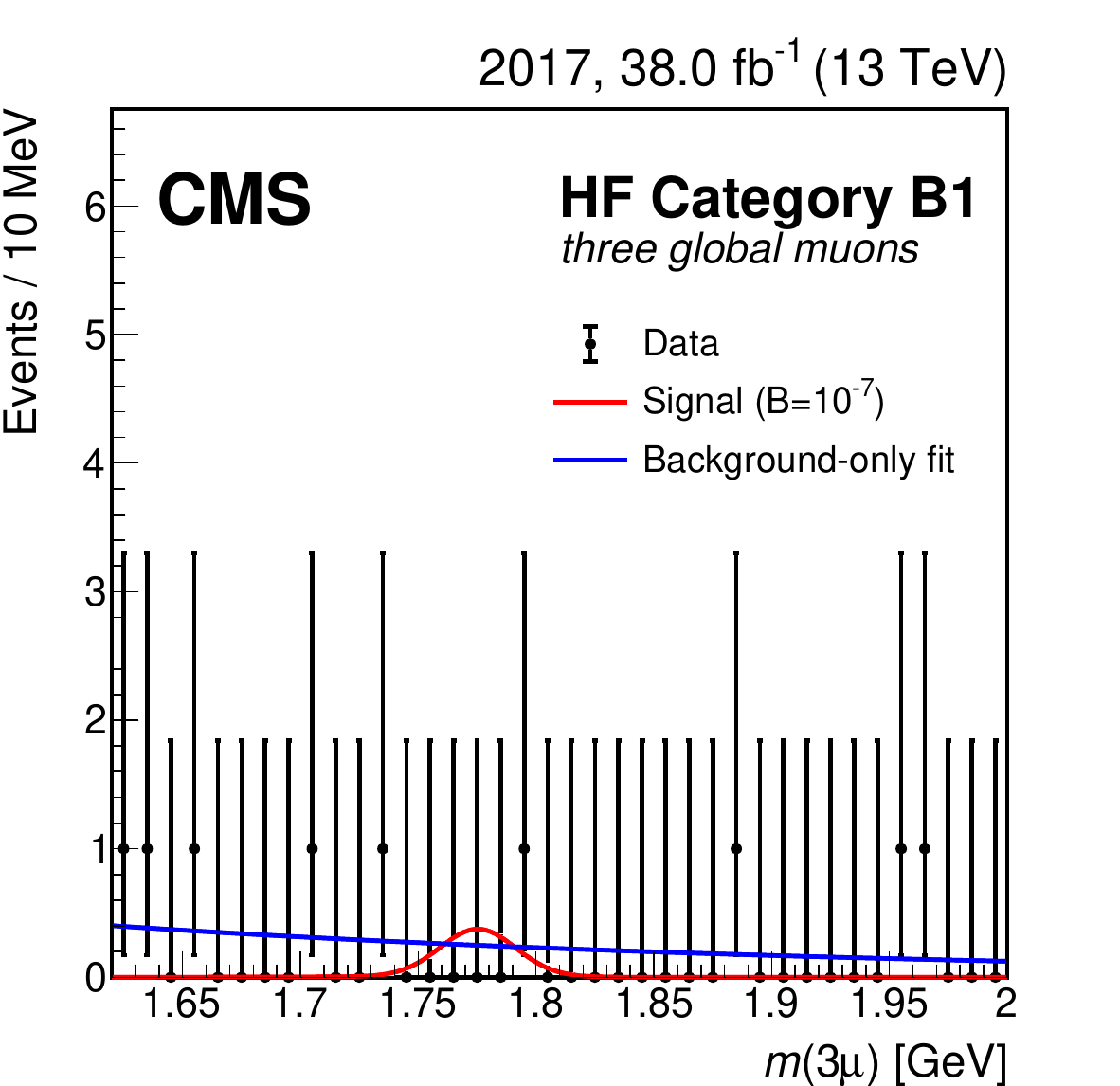}
    \includegraphics[width=\figsixwidth]{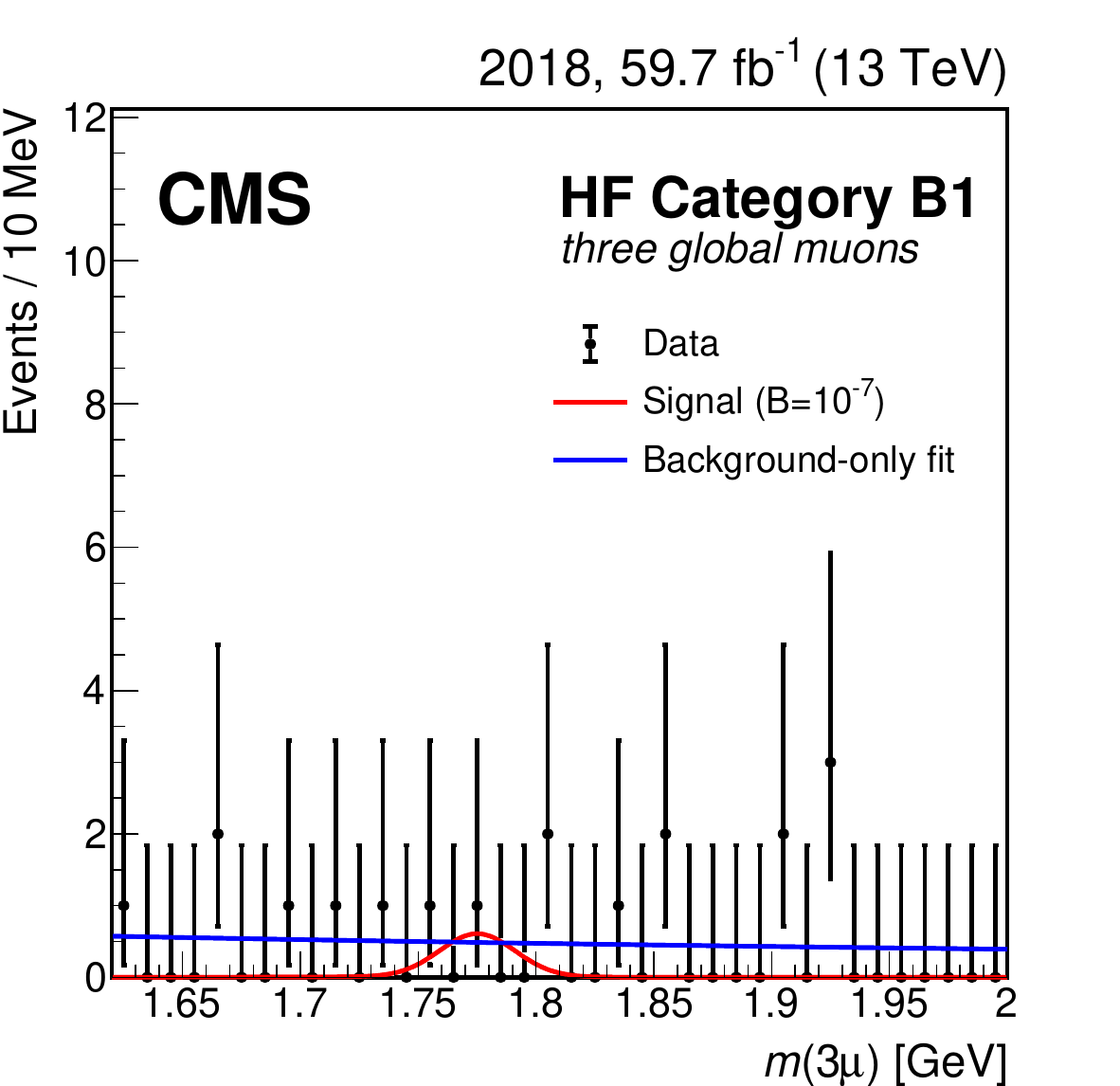}
    \includegraphics[width=\figsixwidth]{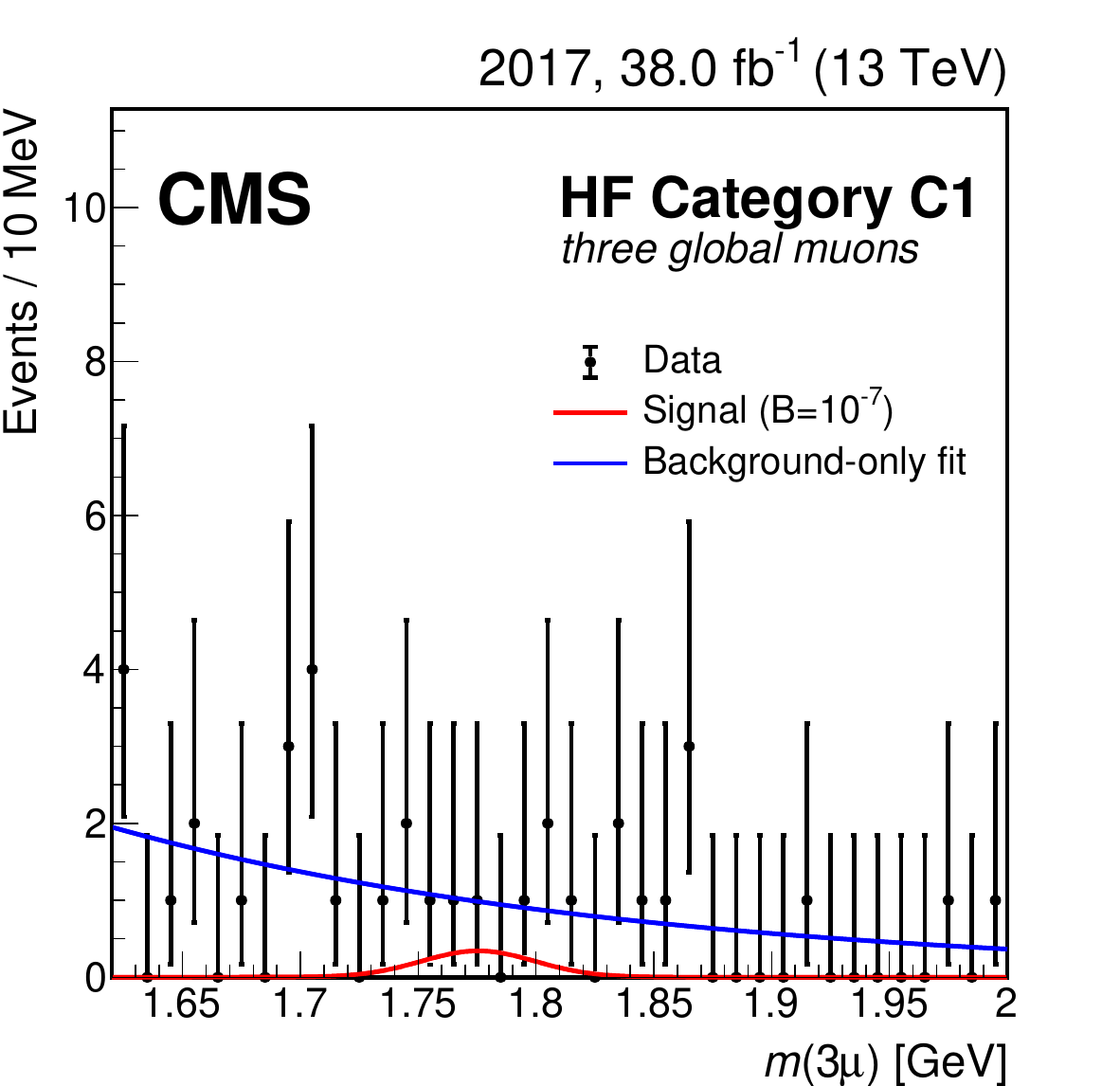}
    \includegraphics[width=\figsixwidth]{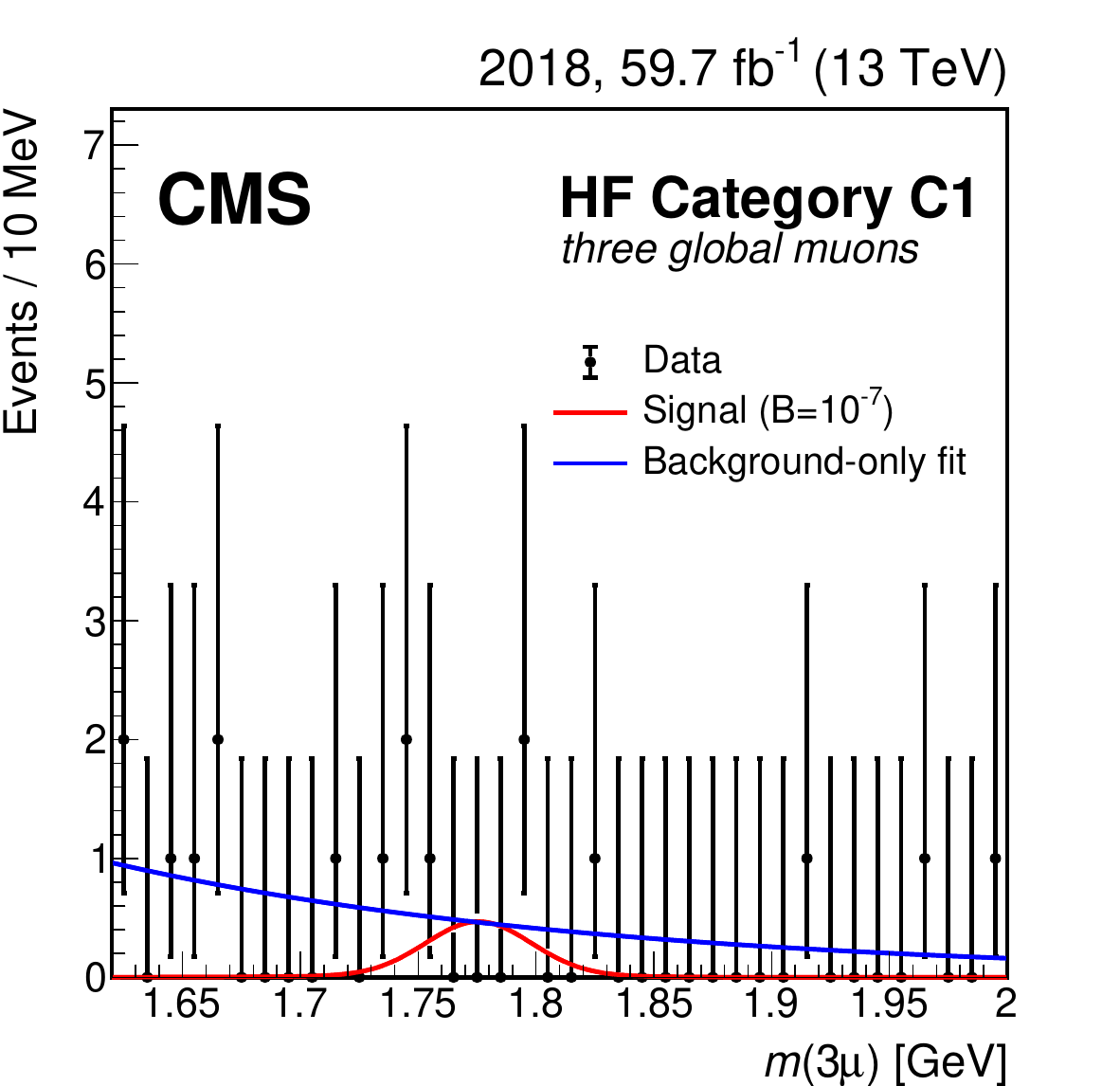}
 \caption{Trimuon mass distributions in the highest BDT score subcategory of each of the three mass resolution
categories of the heavy-flavor (HF) analysis: A1 (upper), B1 (middle), and C1 (lower) for 2017 (left) and 2018 (right) candidate
events with three global muons. Data are shown with black markers. The background-only fit and the expected
signal for $\mathcal{B}(\tautrimu) = 10^{-7}$ are shown with blue and red lines, respectively.
}
\label{fig:hf_3glb}
\end{figure*}

The trimuon invariant mass distributions for categories A1, B1, and C1 are shown in Fig.~\ref{fig:hf_3glb},
along with a background-only fit with an exponential function, and the contribution expected from a signal with
$\mathcal{B}(\tautrimu)$ set to $10^{-7}$, which is chosen to enhance the visibility of the signal in the figures.

\begin{figure*}[hbtp]
  \centering
    \includegraphics[width=\figsixwidth]{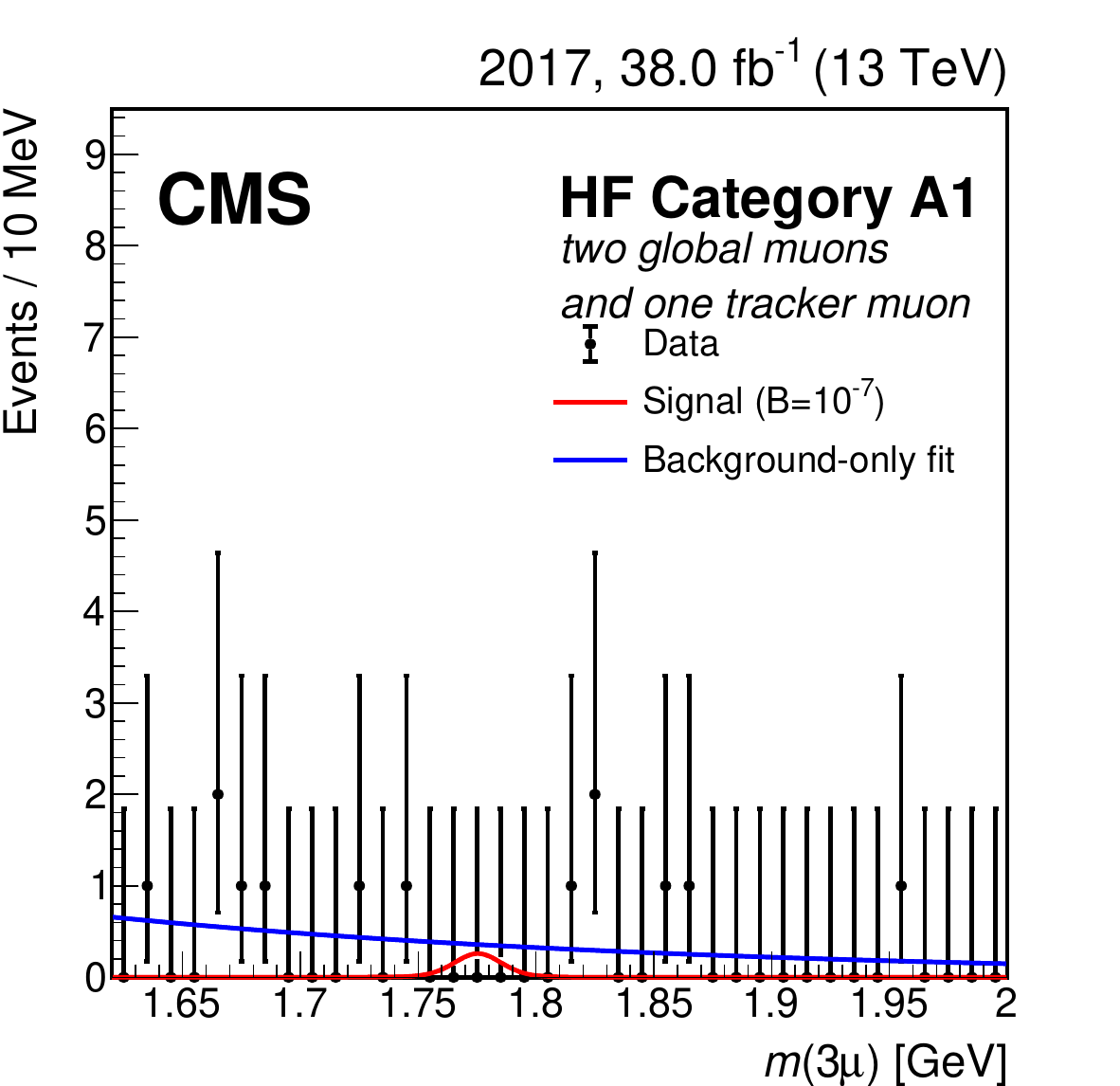}
    \includegraphics[width=\figsixwidth]{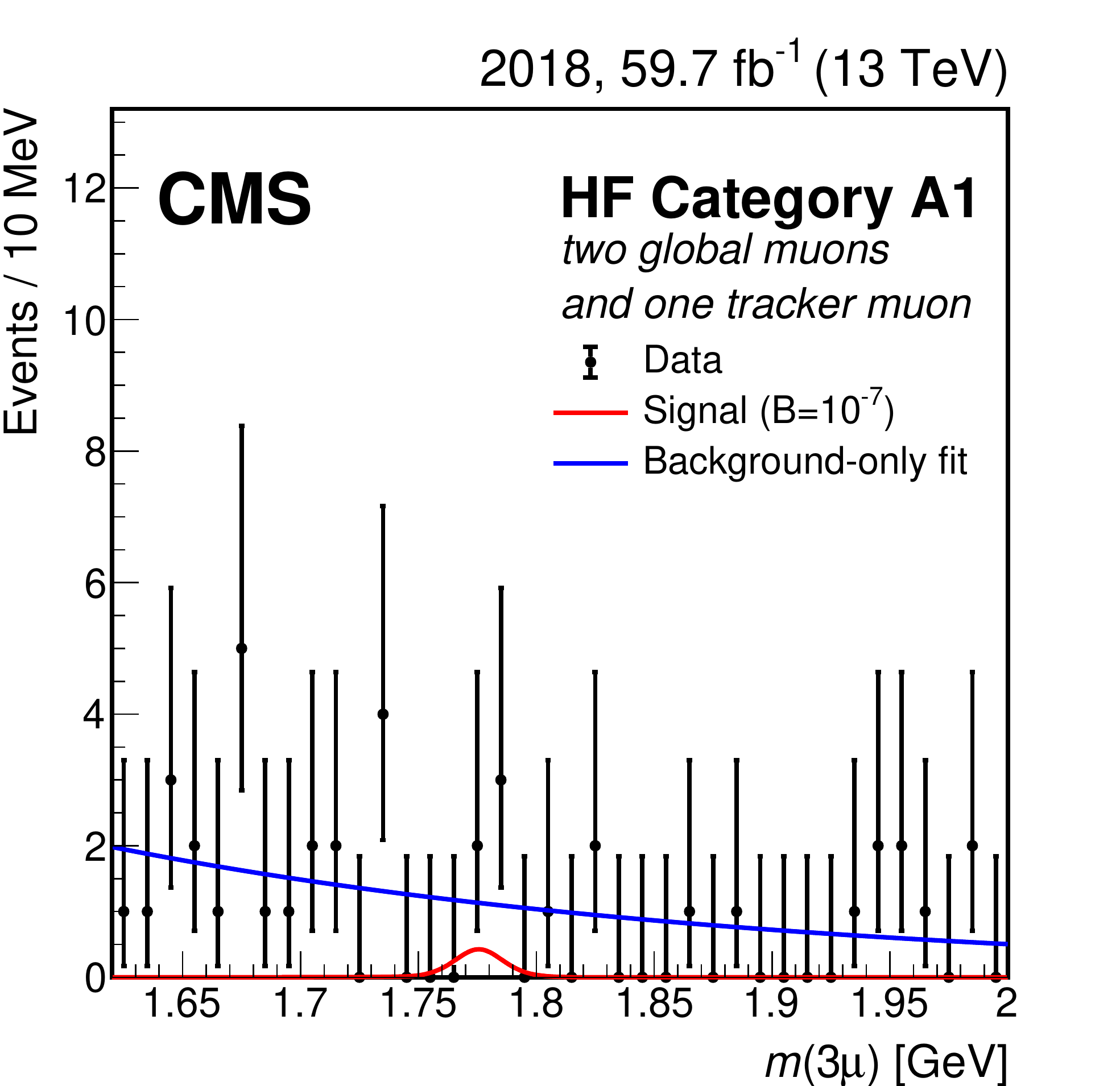}
    \includegraphics[width=\figsixwidth]{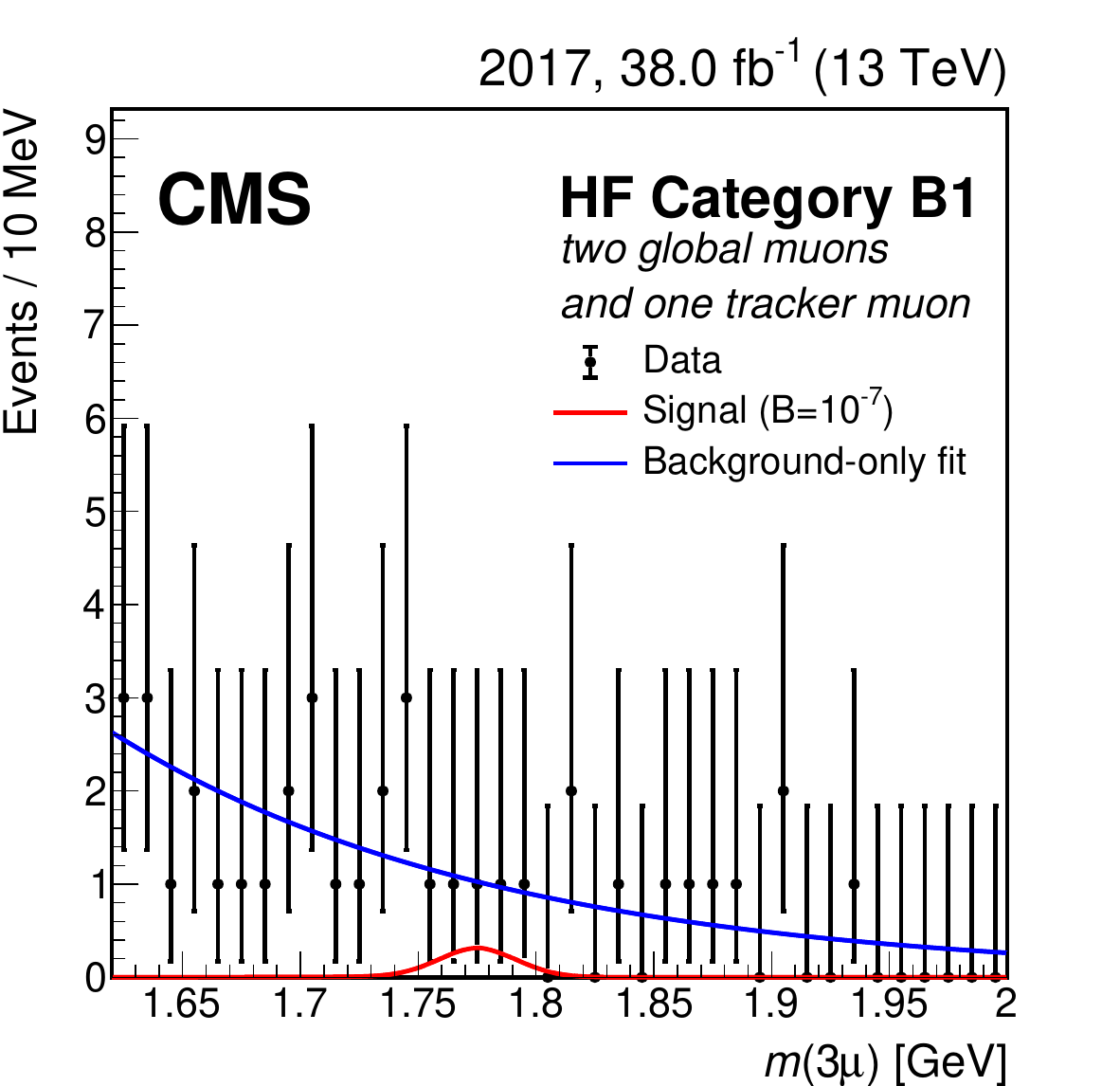}
    \includegraphics[width=\figsixwidth]{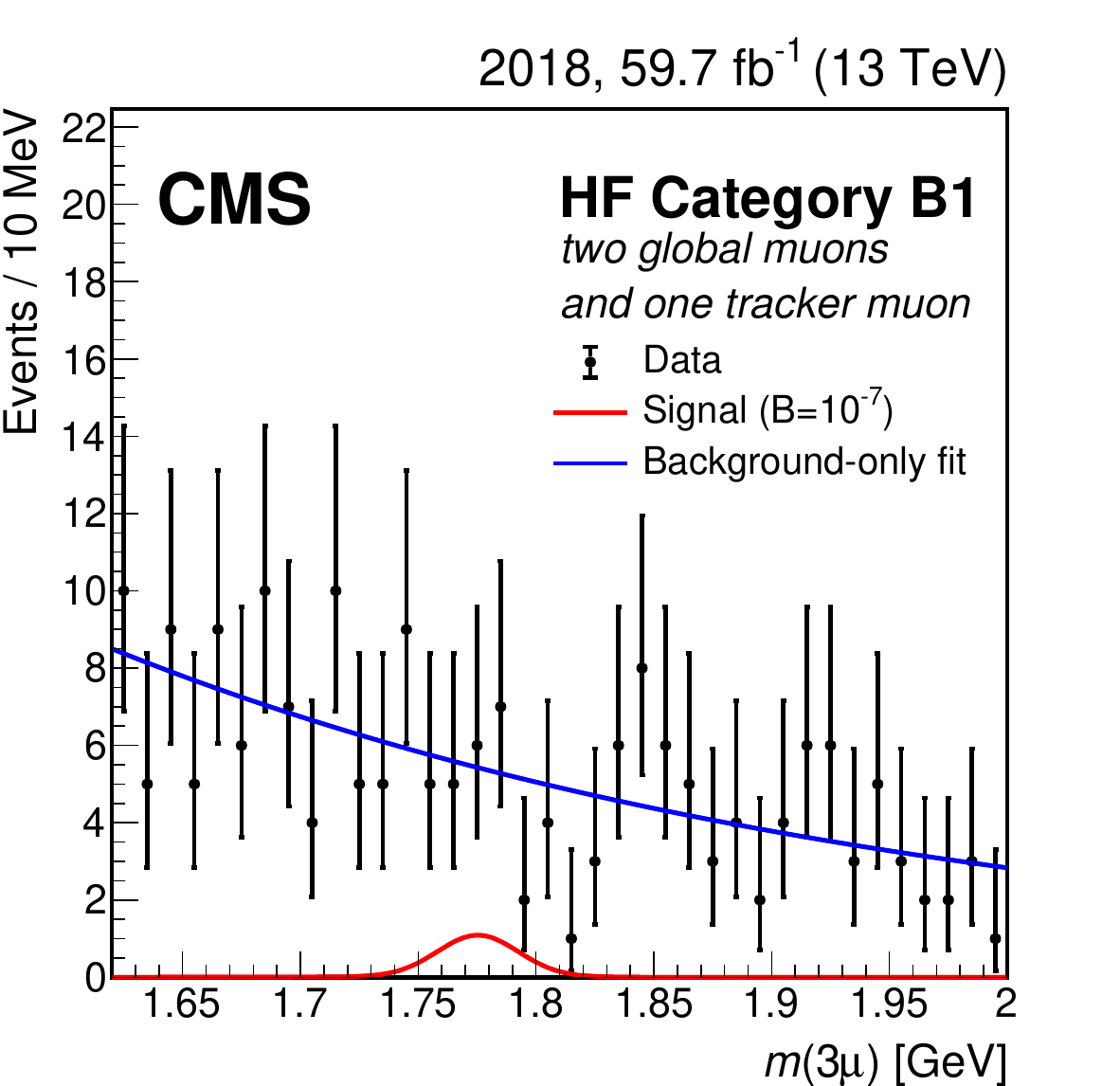}
    \includegraphics[width=\figsixwidth]{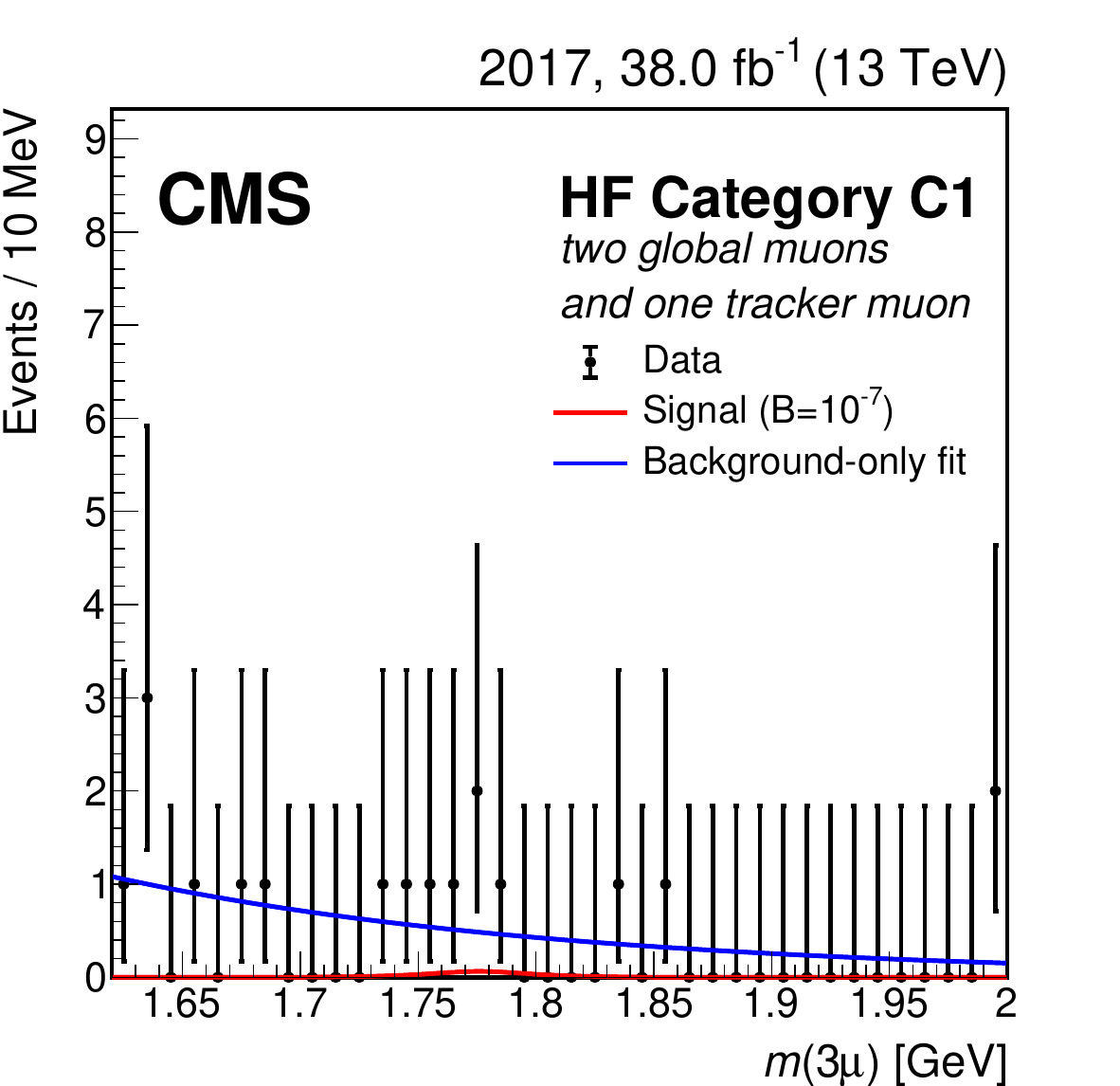}
    \includegraphics[width=\figsixwidth]{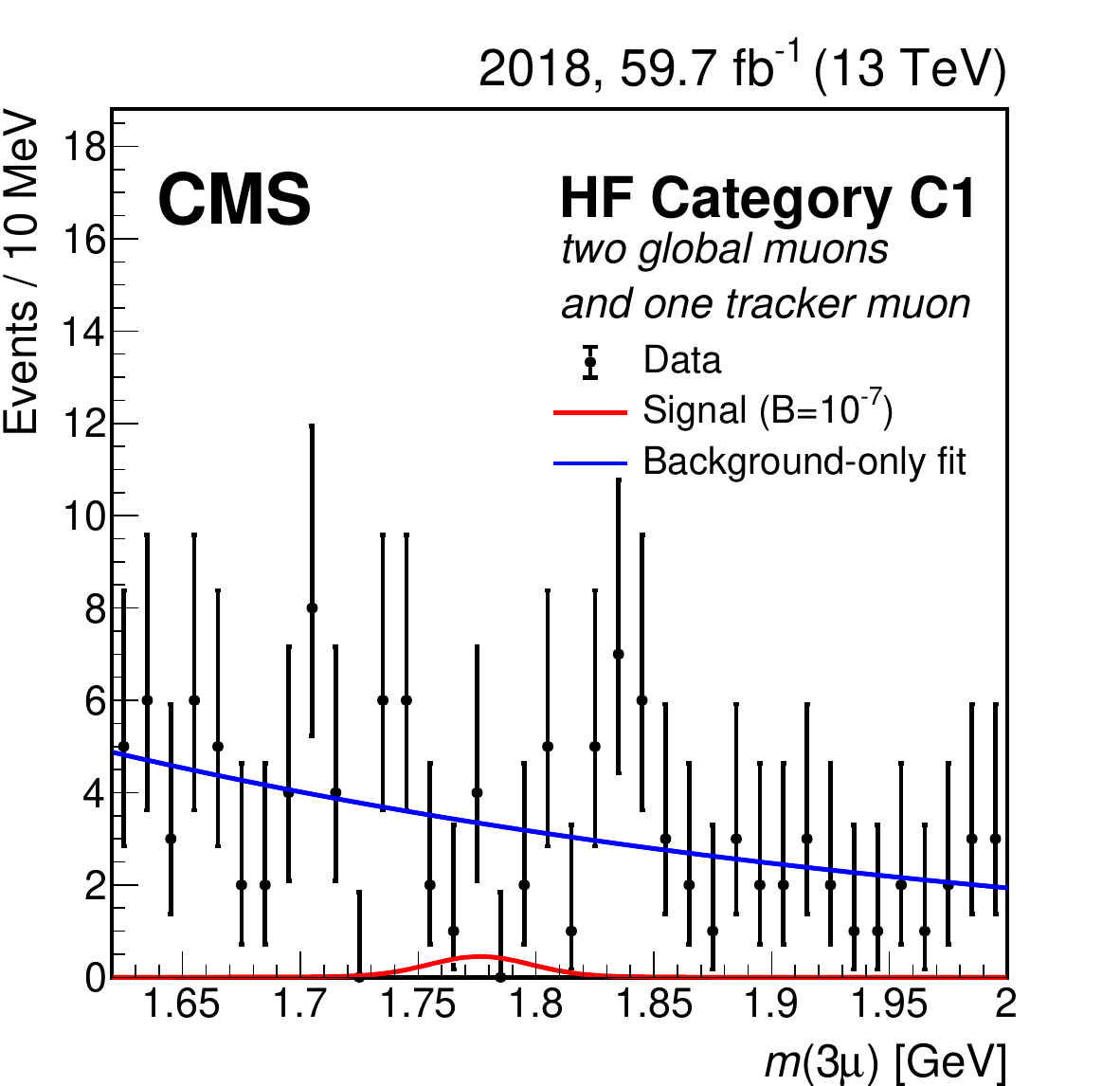}
 \caption{ Trimuon mass distributions in the highest BDT score subcategories of each of the three mass resolution
categories of the heavy-flavor (HF) analysis: A1 (upper), B1 (middle), and C1 (lower) for 2017 (left) and 2018 (right) candidate
events with two global muons and one tracker muon. Data are shown with black markers. The background-only fit and the expected
signal for $\mathcal{B}(\tautrimu) = 10^{-7}$ are shown with blue and red lines, respectively.
}
\label{fig:hf_2glb}
\end{figure*}

Events that do not have a \PGt candidate formed by three global muons but by two global muons
 and one tracker muon~\cite{Sirunyan:2018fpa},
 which is reconstructed as a silicon tracker track extrapolated to match stubs in the muon detectors,
are analyzed separately. The event selection is the same as in the case of three global muons.
A tracker muon reconstruction quality BDT is trained to suppress misidentified muons. Events are then categorized based on
mass resolution and using another analysis BDT,
in the same way as in the case of three global muons, but defining three subcategories instead of four, with the lowest
analysis BDT score subcategory discarded.
Thus six event categories are defined for each year.
The trimuon invariant mass distributions for these two global muons and one tracker muon analysis
categories A1, B1, and C1 are shown in Fig.~\ref{fig:hf_2glb},
along with a background-only fit, with an exponential function, and the contribution expected for a signal of
$\mathcal{B}(\tautrimu)$ set to $10^{-7}$.

The dominant systematic uncertainty is related to the signal normalization, including
the statistical uncertainty in the yield of \dsphipi decays and its variation over different data-taking periods;
the uncertainties in various heavy-flavor decay branching fractions~\cite{PDG2022}
that affect the signal normalization;
the uncertainty in the muon reconstruction efficiency, measured using a tag-and-probe method~\cite{CMS:2011aa} applied to $\cPJgy \!\to\! \mumu$ data events;
and the uncertainty in the BDT requirement efficiency, studied by training a BDT for the \dsphipi
process using the same observables and comparing the \dsphipi efficiencies as functions of the
BDT requirement in data and simulated events.
The uncertainties in the mean and width of the signal trimuon invariant mass distribution are determined by comparing
\dsphipi shapes in data and simulated events in each mass resolution category.
The trimuon mass distribution of the background in each event category can be modeled analytically
using an exponential, a power-law, or a second-order polynomial function. The systematic uncertainty
associated with the choice of the function is treated as a discrete nuisance parameter in the
fit~\cite{Dauncey:2014xga}.

\section{Search for the \texorpdfstring{\tautrimu}{tau to 3mu} decay in \texorpdfstring{\PW}{W} boson events}
\label{sec:Wsearch}

The $\PWp\to\taunu\to\mumu\munut$ signal features three collimated
muons with relatively high \pt.
The trimuon system is isolated from any hadronic activity in the event and associated with large
missing momentum carried away by the neutrino.

A dedicated HLT path was used in 2017--2018 in the search for the \tautrimu decay in \PW boson events
collected with the same L1 trigger requirements as those described in Section~\ref{sec:HFsearch}, 
as well as events collected with L1 trigger criteria targeting single-muon events with a muon \pt greater than 22 (25)\GeV in 2017 (2018). 
Three muons were required to be reconstructed at the HLT, each with $\pt>1$\GeV, 
and at least one with $\pt>7$\GeV.
The \pt of the trimuon system was required to be greater than 15\GeV and the resulting \PGt candidate
was required to have an invariant mass between 1.3 and 2.1\GeV.
The three muons were required to have a total charge of $\pm$1.
An isolation variable, defined as the \pt sum of all tracks, other than those associated with the three muons,
with $\Delta R < 0.8$ with respect to the \PGt candidate, and with a
distance of closest approach with respect to the \PGt candidate below 3\mm along the beam direction,
was required to be smaller than 20\% of the \PGt candidate \pt.

The offline selection requires three muons with $\pt>3.5\,(2.0)\GeV$ for $\abs{\eta} < 1.2$ ($1.2 < \abs{\eta} < 2.4$),
with the three muons matching those that triggered the event.
Tau lepton candidates are formed by combining all possible triplets of muons that pass the selection and can be successfully fitted to a common vertex.
To remove contamination from dimuon resonances, a \PGt candidate is discarded if
the mass of any pair of oppositely charged muons belonging to the \PGt candidate lies within 20\MeV of the \PGf or the \PGoP{782} masses.
No more than one \PGt candidate per event is considered. Preference is given to candidates
with the largest transverse mass
 $\mT = \sqrt{\smash[b]{2\pt^{\vphantom{a}\smash{\PGt}}\ptmiss[1 - \cos{\Delta\phi(\ptvec^{\vphantom{a}\smash{\kern1pt\PGt}}, \ptvecmiss)}]}}$,
which is one of the most sensitive observables to distinguish the \PW boson decay from other processes.
The tracks of the three muon candidates are then refitted using the common vertex constraint.

The expected number of \tautrimu signal events from \PW boson decays, denoted as $N_{3\PGm(\PW)}$, is related to $\mathcal{B}(\tautrimu)$ by
\ifthenelse{\boolean{cms@external}}
{
\begin{linenomath}
\begin{equation*}
\begin{aligned}
N_{3\PGm(\PW)} \, &= \,
\mathcal{L}\, \sigma(\ppWX) \, \mathcal{B}(\Wtaunu)\\
& \times \,
\mathcal{A}_{3\PGm(\PW)}\, \epsilon_{3\PGm(\PW)}
\, \mathcal{B}(\tautrimu),
\end{aligned}
\end{equation*}
\end{linenomath}
}
{
\begin{linenomath}
\begin{equation*}
N_{3\PGm(\PW)} \, = \,
\mathcal{L}\, \sigma(\ppWX) \, \mathcal{B}(\Wtaunu)\,
\mathcal{A}_{3\PGm(\PW)}\, \epsilon_{3\PGm(\PW)}
\, \mathcal{B}(\tautrimu),
\end{equation*}
\end{linenomath}
}
where $\mathcal{L}$ is the integrated luminosity~\cite{CMS:2021xjt,CMS:2018elu,CMS:2019jhq},
$\sigma(\ppWX)$ is the \PW boson production cross section, $\mathcal{B}(\Wtaunu)$ is the branching
fraction of the \PW boson decay to $\PGt\PGnGt$, $\mathcal{A}_{3\PGm(\PW)}$ is the acceptance, and $\epsilon_{3\PGm(\PW)}$ is
the combined trigger and selection efficiency for the three muons.  The product of $\sigma(\ppWX)$ and
$\mathcal{B}(\Wtaunu)$ is obtained from the ATLAS measurement of
$\sigma(\ppWX) \mathcal{B}(\Wmunu)$ at 13 TeV~\cite{Aad:2016naf} and the world-average value of the ratio
$\mathcal{B}(\Wtaunu) / \mathcal{B}(\Wmunu)$~\cite{PDG2022}.
The \pt and $\eta$ distributions of the \PW boson sample generated at LO are reweighted to match the distributions obtained from
the smaller sample simulated at NLO precision.

\begin{figure*}[hbtp]
  \centering 
    \includegraphics[width=0.48\textwidth]{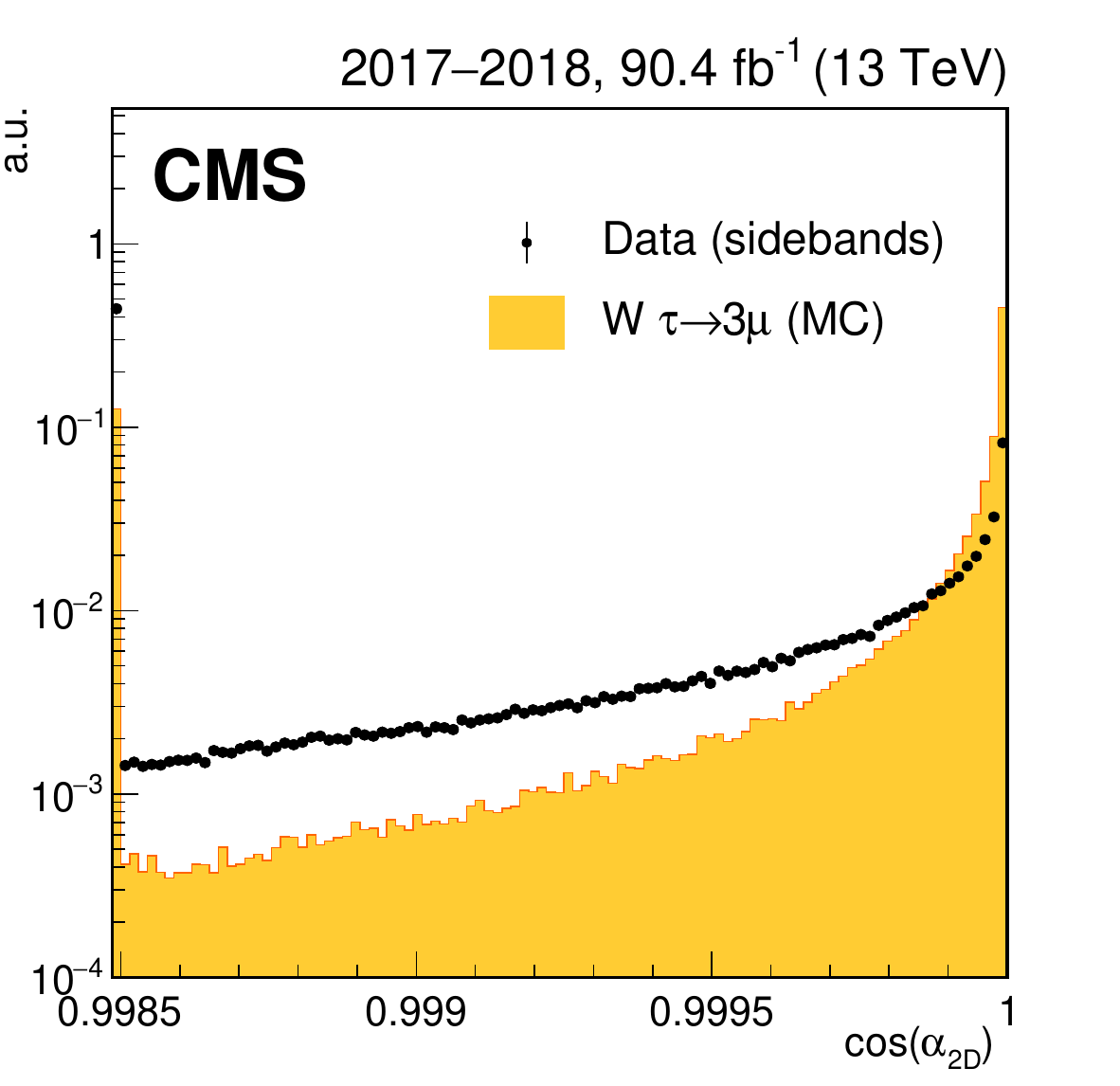}
    \includegraphics[width=0.48\textwidth]{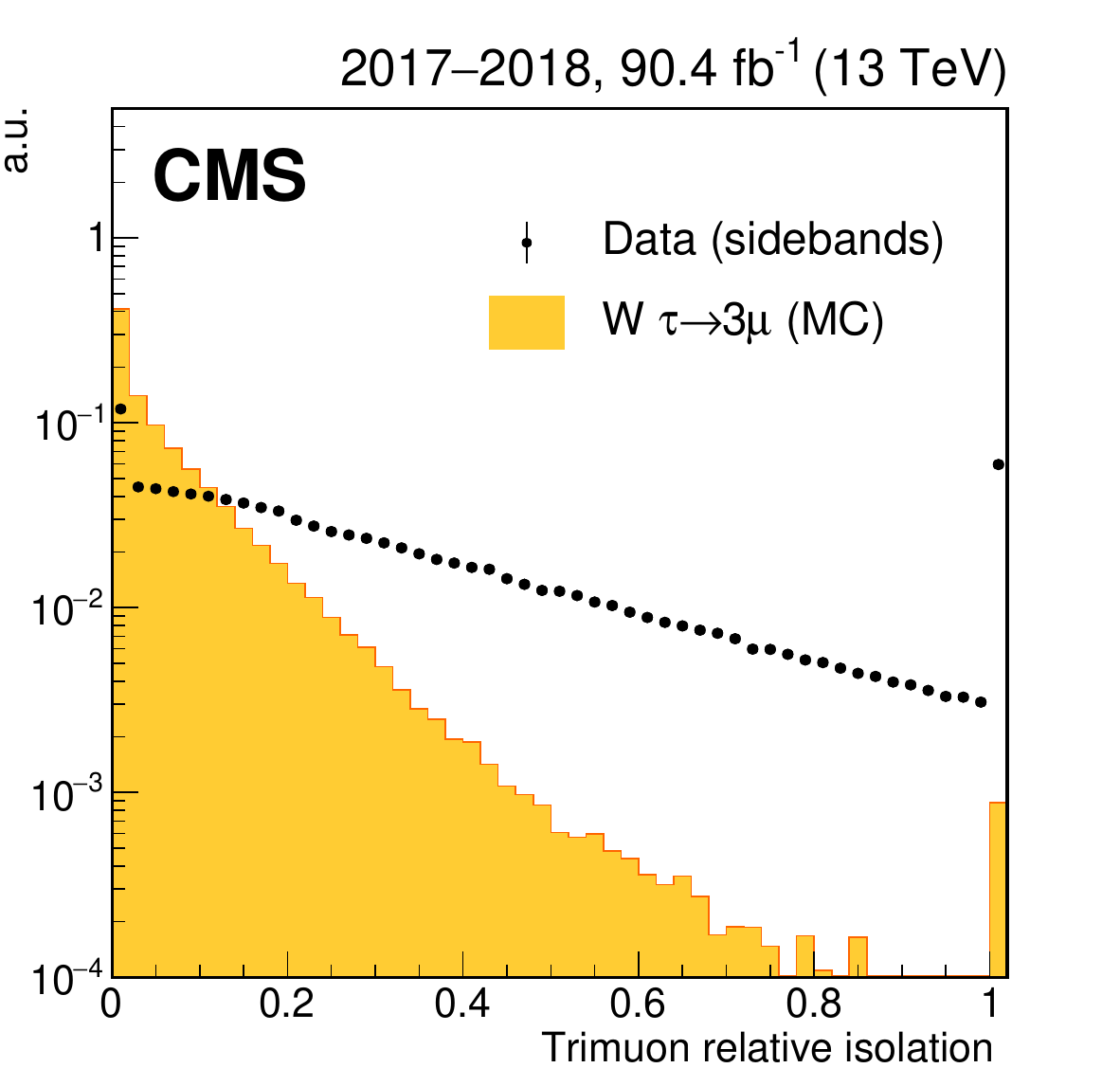}
    \includegraphics[width=0.48\textwidth]{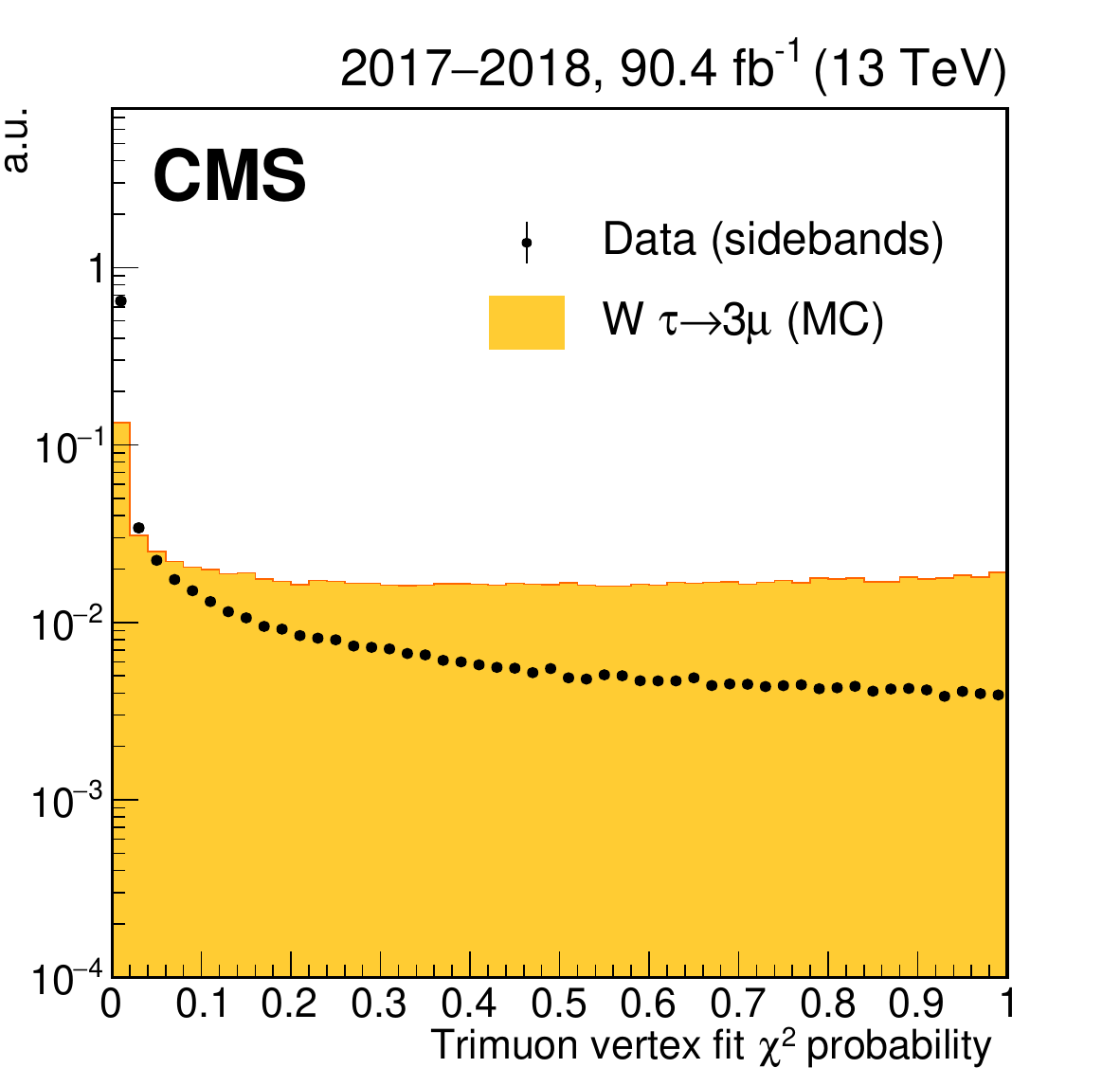}
    \includegraphics[width=0.48\textwidth]{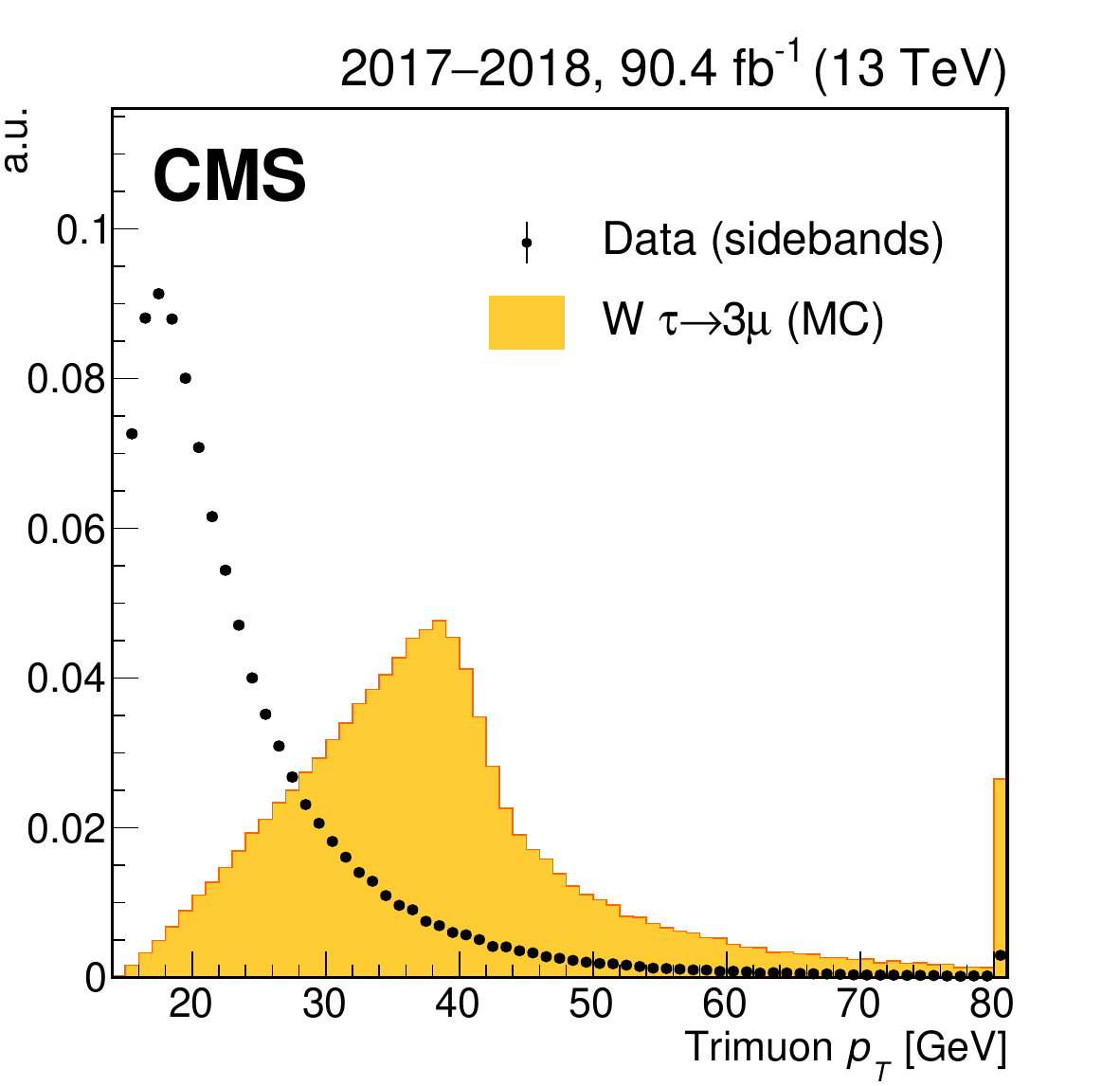}
 \caption{ Signal and background distributions for the four observables with the highest discrimination
power used for the \PW boson analysis BDT training: cos($\alpha_{\mathrm{2D}}$) (upper left),
relative isolation observable of the trimuon system (upper right),
trimuon vertex fit $\chi^2$ probability (lower left),
trimuon \pt (lower right).
The signal and background distributions are obtained respectively from MC simulation
and from the mass-sideband regions in data.
All distributions are normalized to unit area.
The leftmost (rightmost) bins include underflow (overflow).
}
\label{fig:W_BDT}
\end{figure*}

Residual backgrounds are further reduced using a BDT, trained on simulated $\PWp\to\taunu\to\mumu\munut$ signal events
and a background sample from data with the trimuon invariant mass within the 1.60--1.74 or 1.82--2.00\GeV range.
The inputs to the BDT can be categorized into three different classes, related to either:
$\PWp\to\taunu$ decay, \PGt decay vertex, or quality of muon reconstruction. 
The $\PWp\to\taunu$ related observables are: the \pt, $\eta$, and relative isolation
of the \PGt candidate, along with $\mT$, \ptmiss and the \PW~boson \pt
(defined as the magnitude of the vector sum of \ptvec of the three muons and \ptvecmiss).
The isolation observable is defined as the scalar \pt sum of photons and tracks with
$\Delta R < 0.8$ around the \PGt candidate direction, excluding the three muon candidates themselves, and is corrected for the pileup contribution. 
The relative isolation value is obtained by dividing the absolute isolation by the \PGt candidate \pt.
In addition, the longitudinal component of the missing momentum is used: this is inferred from solving the energy-momentum equation for the \PGt candidate and the \ptvecmiss after imposing the nominal \PW boson mass~\cite{PDG2022} constraint.
Observables related to the \PGt decay vertex are the $\chi^2$ of the vertex fit, the significance of the distance in the transverse plane between the beamline and the decay vertex, and $\alpha_{\mathrm{2D}}$.
The quality of muon reconstruction input is a single bit for each muon indicating whether the muon
has passed the tight muon identification requirement~\cite{Sirunyan:2018fpa}.
The distributions of the observables providing the largest discrimination power between signal and background are shown in
Fig.~\ref{fig:W_BDT}.

The events are separated into categories A, B, and C, in the same way as for the heavy-flavor analysis,
from higher to lower mass resolutions.
For each mass resolution category, an optimal BDT cut value is chosen, with events below the cut
value rejected.
It is found that optimizing for the largest expected signal significance or
the most stringent expected upper limit has very little effect on the cut value obtained.
The signal efficiency and background rejection of the BDT are checked to be uniform in \PGt candidate mass
to avoid bias.

All selection criteria described above were determined in a blind way, without looking at data
events with trimuon invariant mass in the range 1.74--1.82\GeV. 
Once the data were unblinded, it was observed
that a significant fraction of events after the BDT selection had the trimuon
vertex not displaced from the beamline,
in spite of the distance in the transverse plane between the beamline and the trimuon vertex
divided by its uncertainty being one of the BDT input features.
These events were found to be consistent with the MC simulation
prediction of $\PWp\to\mumu\munu$ events, where a \PW~boson decays to a muon
and a neutrino, and the emission of final-state radiation from the muon results in the production of two additional muons. 
To reduce this background, the distance in the transverse plane between the beamline and the trimuon
vertex is required to be larger than twice its uncertainty. 

\begin{figure*}[hbtp]
  \centering
    \includegraphics[width=\figsixwidth]{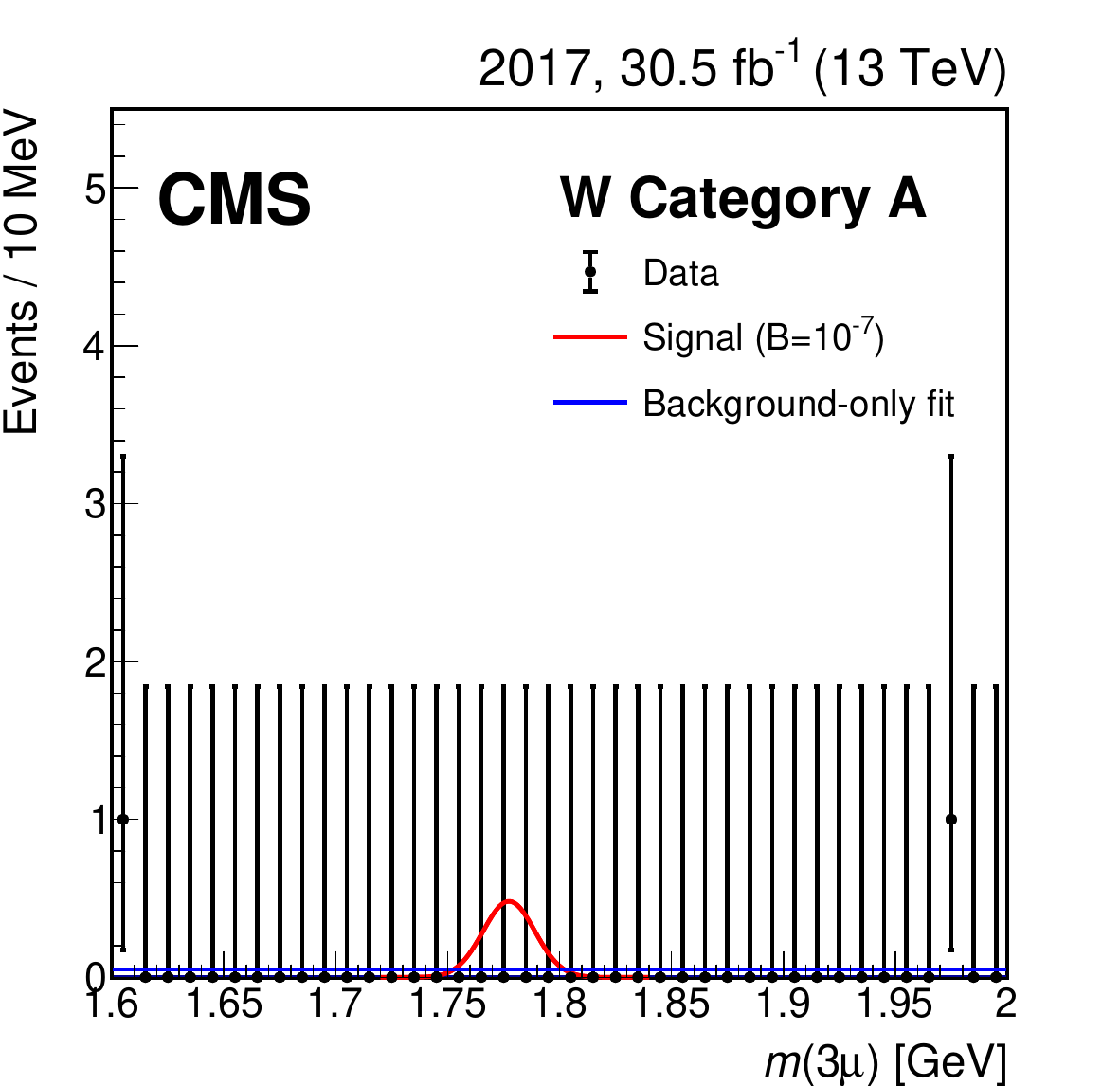}
    \includegraphics[width=\figsixwidth]{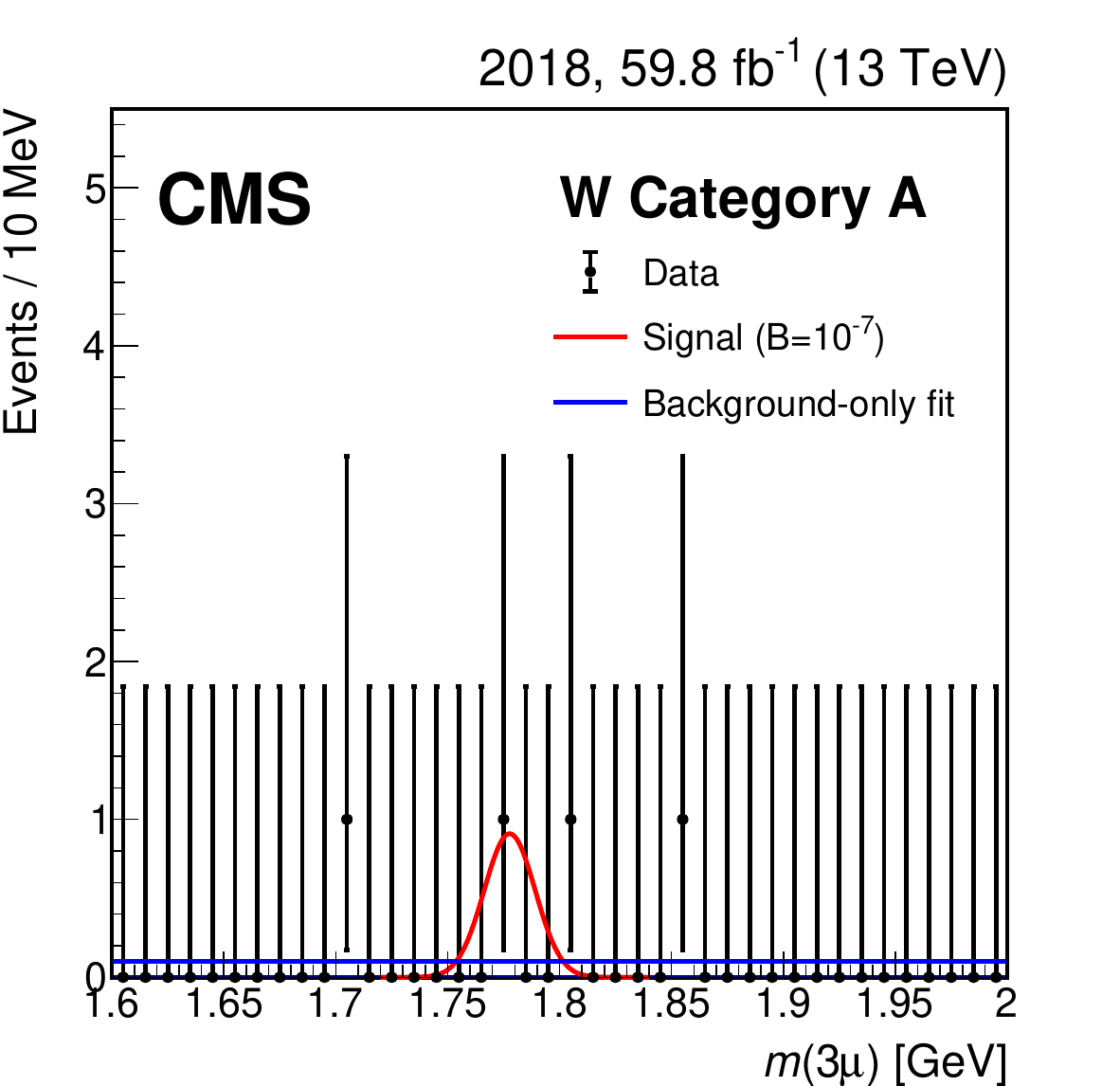}
    \includegraphics[width=\figsixwidth]{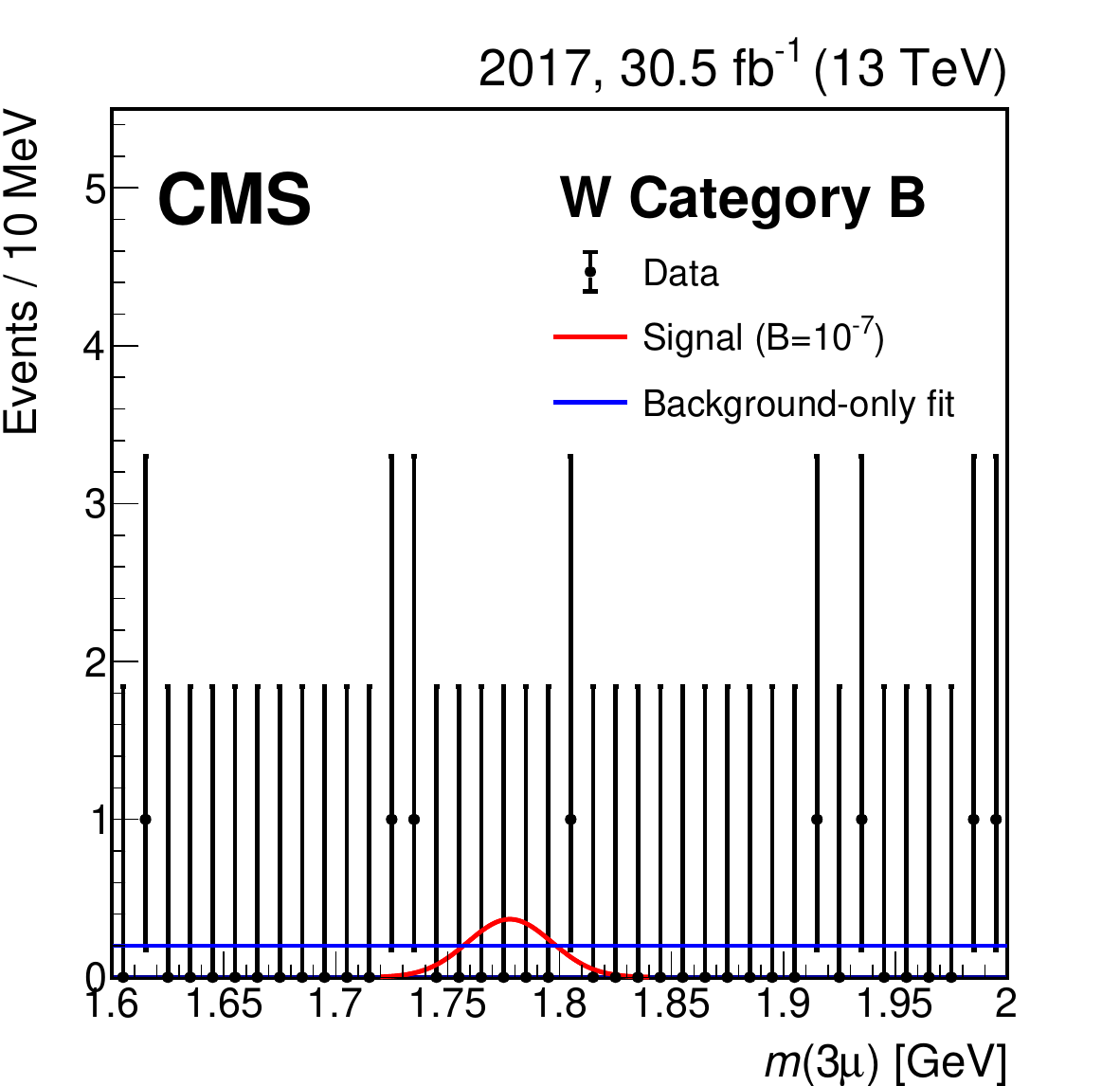}
    \includegraphics[width=\figsixwidth]{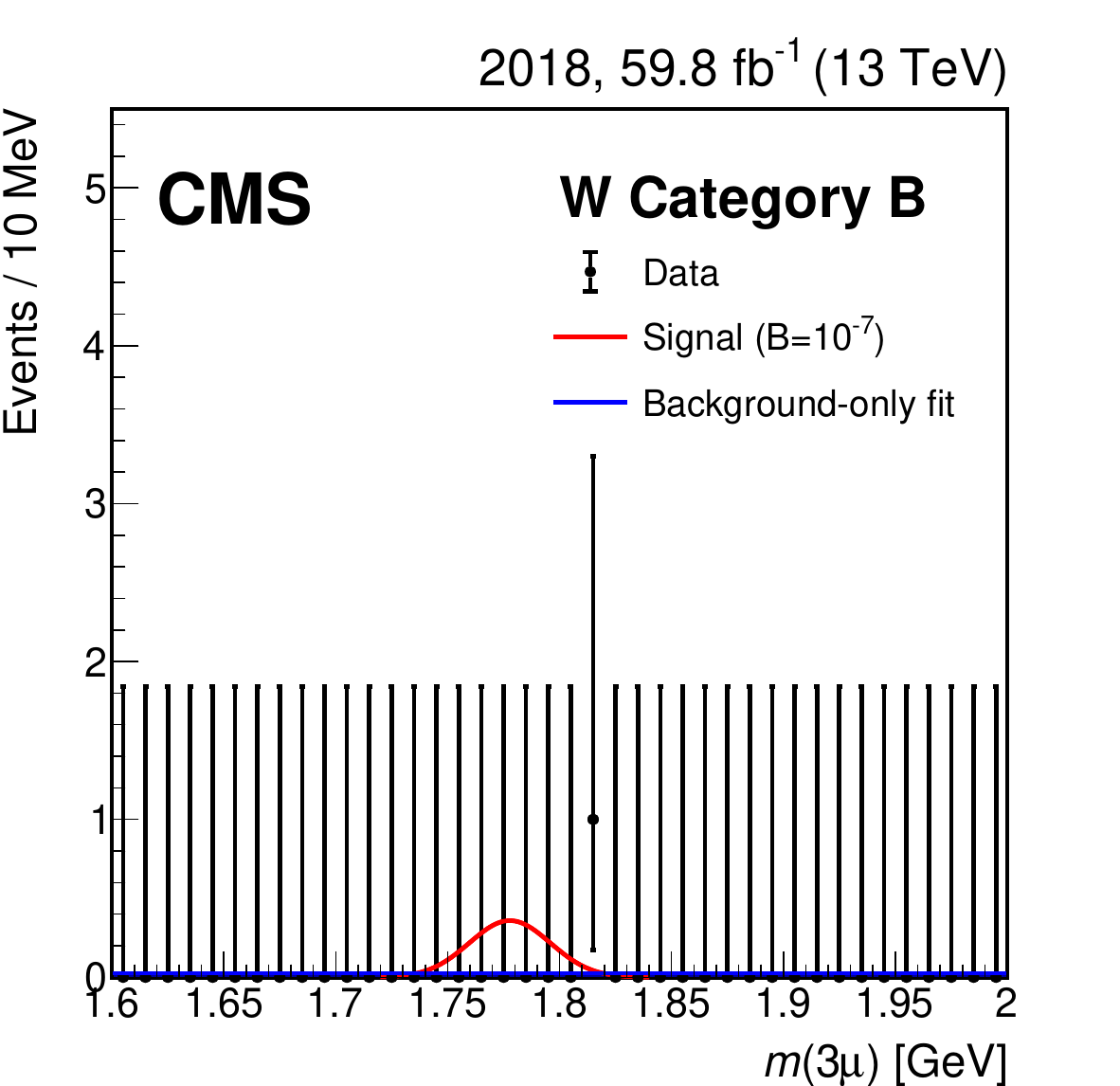}
    \includegraphics[width=\figsixwidth]{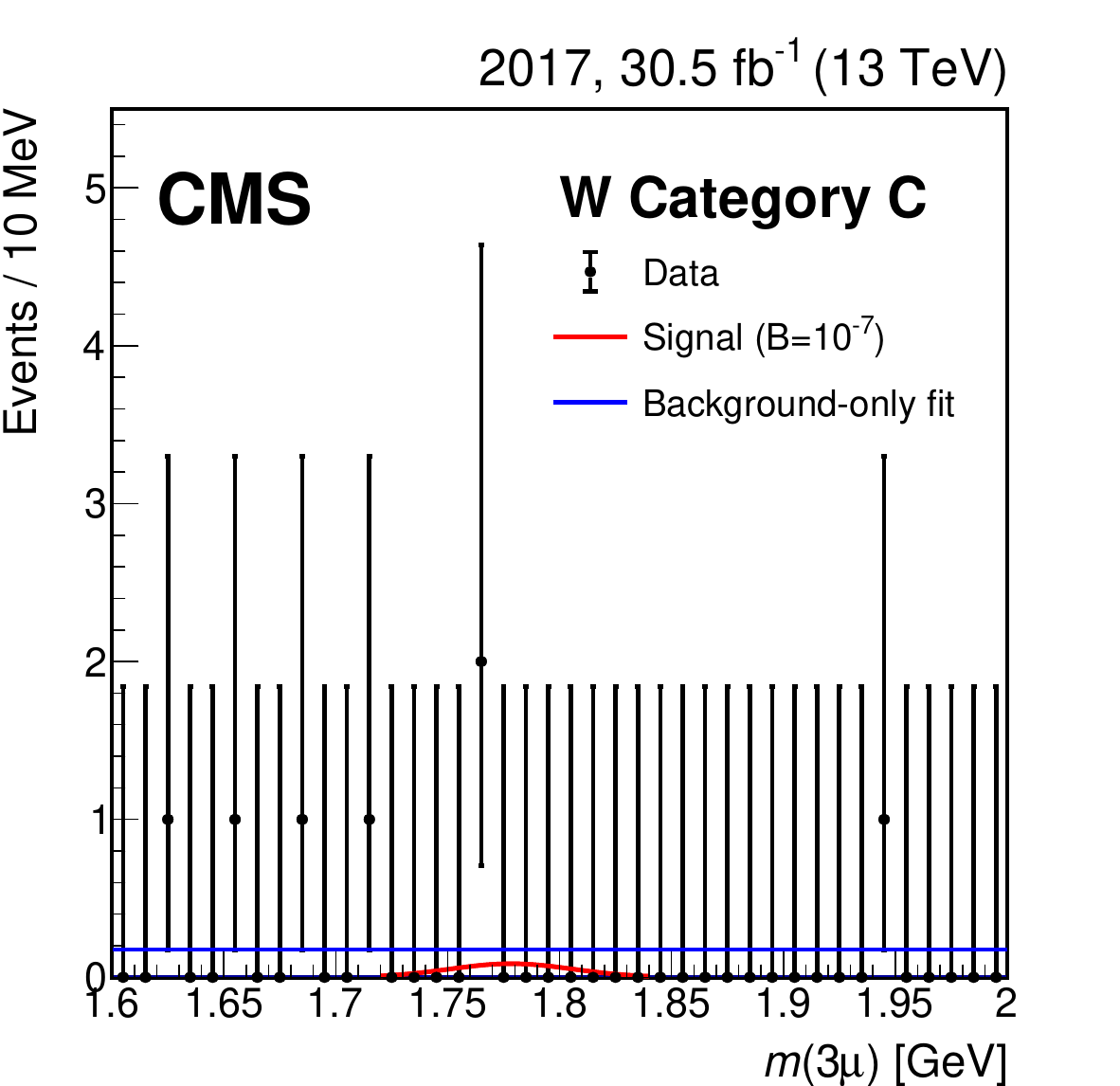}
    \includegraphics[width=\figsixwidth]{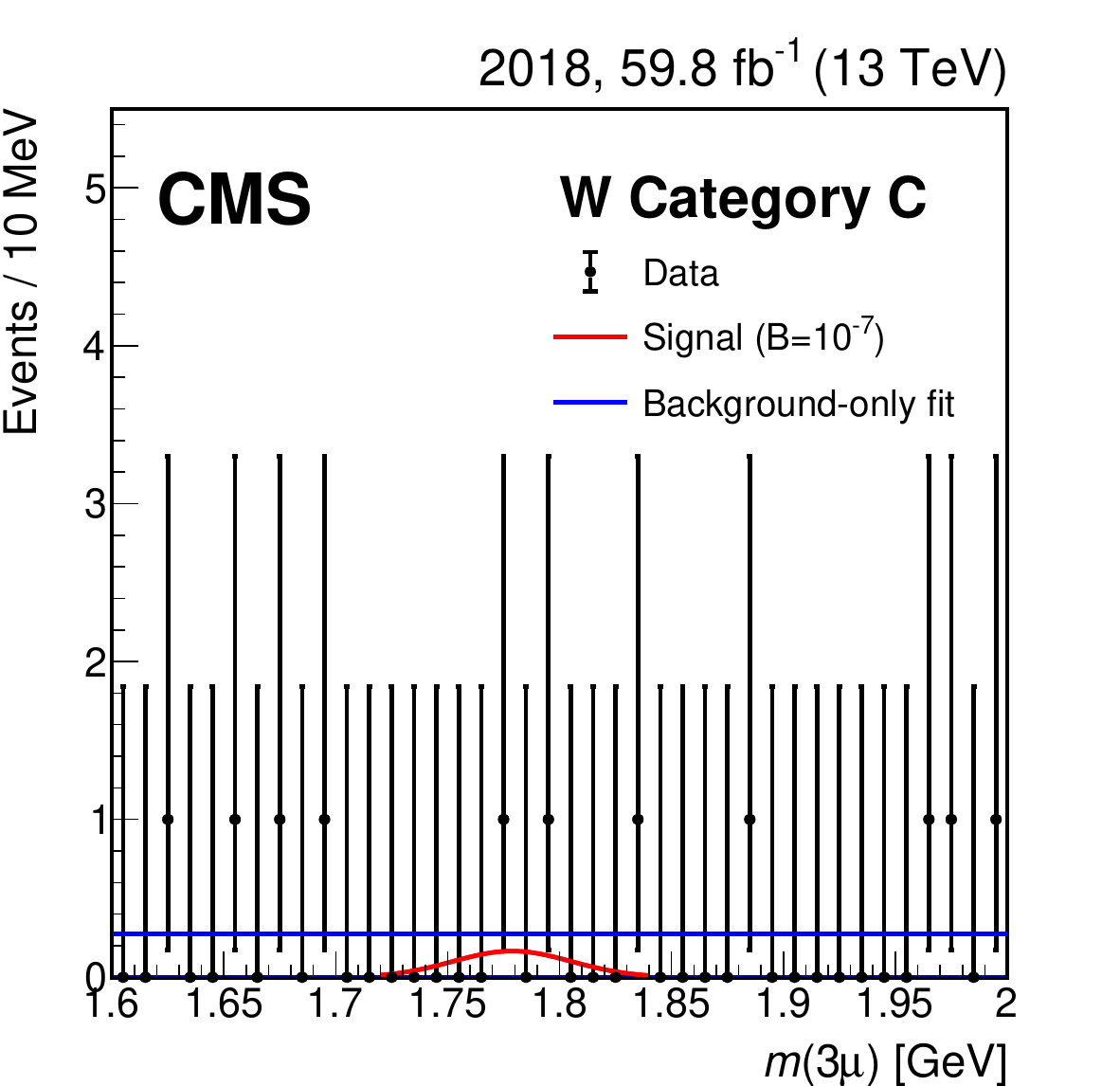}
 \caption{ Trimuon mass distributions of the 2017 (left) and 2018 (right) data events in the three mass resolution categories
A (upper), B (middle), and C (lower) of the \PW boson analysis. Data are shown with black markers. The background-only fit and
the expected signal for $\mathcal{B}(\tautrimu) = 10^{-7}$ are shown with blue and red lines, respectively.
}
\label{fig:w_m3mu}
\end{figure*}

The trimuon invariant mass distributions, after final selections, in categories A, B, and C are shown in Fig.~\ref{fig:w_m3mu},
along with a background-only fit with a flat function, and the contribution expected for a signal of
$\mathcal{B}(\tautrimu)$ set to $10^{-7}$.

The major systematic uncertainties include
the uncertainties in single-muon reconstruction efficiencies, measured using the tag-and-probe method applied to
$\cPJgy \!\to\! \mumu$ data events;
the uncertainty in the HLT isolation selection efficiency, studied by comparing the trigger efficiency for \dsphipi events in data
and simulation selected by an alternative version of the HLT path that contains no isolation requirements;
the uncertainty in the NLO reweighting, which arises from the statistical uncertainty in the NLO simulation.
The uncertainties in the mean and width of the signal trimuon invariant mass distribution are determined
in the same way as in the heavy-flavor analysis.

\section{Results}
\label{sec:results}

The branching fraction $\mathcal{B}(\tautrimu)$ is obtained from a simultaneous unbinned maximum
likelihood fit to the trimuon mass distributions in a range 1.6--2.0\GeV in all 36 event categories of the heavy-flavor
analysis and the \PW boson analysis.
For the heavy-flavor analysis, the signal model is a Gaussian plus Crystal Ball function~\cite{CrystalBallRef},
while for the \PW boson analysis it is a Gaussian function.
Both signal models have fixed mean and width, as determined from fitting the simulated events in the corresponding category.
The background normalizations are free parameters in the fit.
 
Upper limits on $\mathcal{B}(\tautrimu)$ are determined using a frequentist method~\cite{CMS-NOTE-2011-005}
based on a modified profile likelihood test statistic and the \CLs criterion~\cite{Read:2002hq,Junk:1999kv}.
Uncertainties are incorporated via nuisance parameters,
and are assumed to be uncorrelated between the heavy-flavor analysis and the \PW boson analysis.
The nuisance parameters for the low-expected-background \PW boson analysis are treated with the
strategy described in Ref.~\cite{COUSINS1992331}.
Events from data and simulation that pass the selection criteria of both analyses are removed
from the heavy-flavor analysis in the combined fit, to benefit from the higher signal-to-background ratio in the W analysis.

The analysis sensitivity is limited by the statistical uncertainty,
while the total impact of the systematic uncertainties is found to be a few percent.

The heavy-flavor analysis results in an observed (expected) upper limit at 90\% \CL on $\mathcal{B}(\tautrimu)$ of $3.4\:(3.6) \times 10^{-8}$.
The \PW boson analysis yields an observed (expected) upper limit at 90\% \CL on $\mathcal{B}(\tautrimu)$ of
$8.0\:(5.6) \times 10^{-8}$.
The combination of the two analyses leads to an observed (expected) upper limit at 90\% \CL on $\mathcal{B}(\tautrimu)$ of $3.1\:(2.7) \times 10^{-8}$.

\begin{figure}[hbtp]
  \centering
    \includegraphics[width=\figwidewidth]{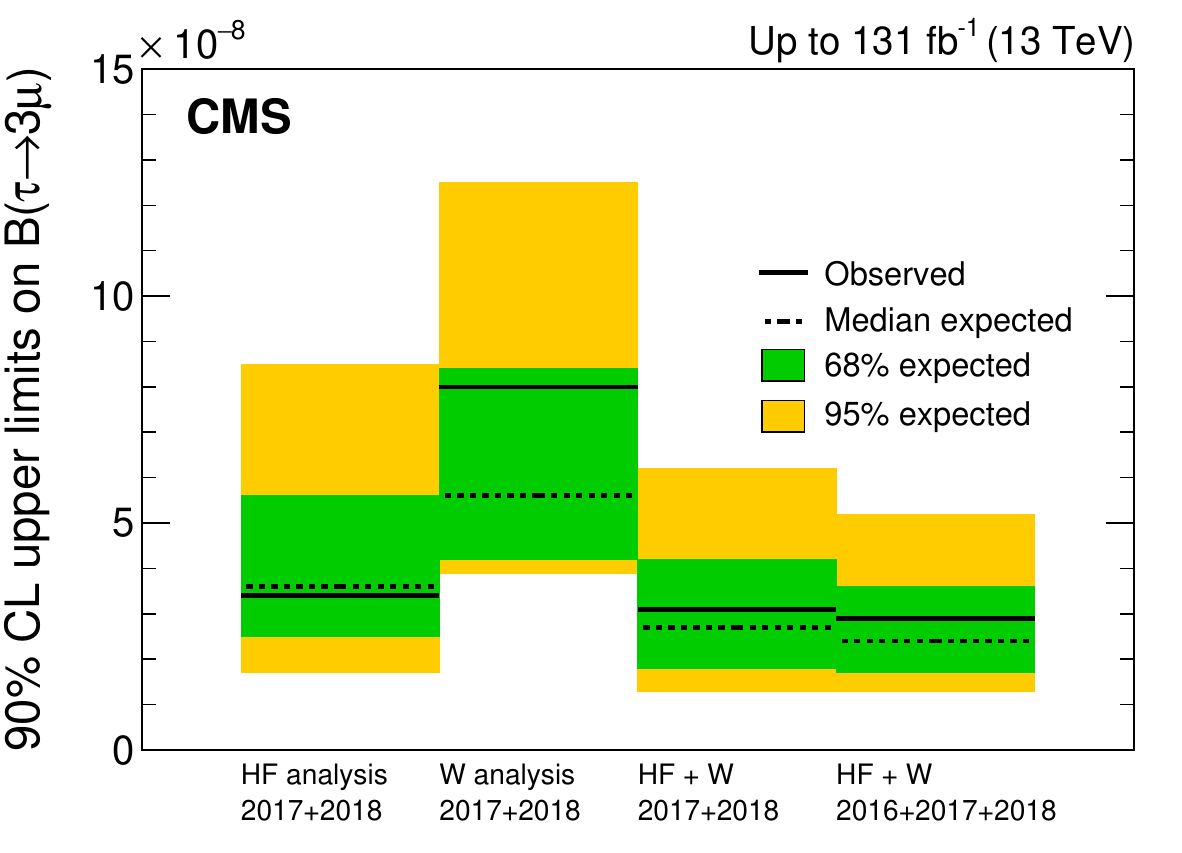}
 \caption{
Observed and expected upper limits on $\mathcal{B}(\tautrimu)$ at 90\% \CL, from the heavy-flavor (HF)
analysis, the \PW boson analysis, the combination of the two analyses, as well as
their combination with the previously published result using 2016 data.
}
\label{fig:upper_limits}
\end{figure}

The previously published result based on 2016 data~\cite{CMS:2020kwy} is combined with the 
new results by performing a simultaneous unbinned maximum
likelihood fit to the trimuon mass distributions,
leading to an observed (expected) upper limit at 90\% \CL on $\mathcal{B}(\tautrimu)$ of
$2.9\:(2.4) \times 10^{-8}$.
The upper limits at 90\% \CL are summarized in Fig.~\ref{fig:upper_limits}.
The observed (expected) upper limit at 95\% \CL is $3.6\:(3.0) \times 10^{-8}$.

\section{Summary}
\label{sec:summary}

A search has been presented for the lepton flavor violating decay \tautrimu, using proton-proton collisions at
a center-of-mass energy of 13\TeV recorded by the CMS experiment at the
LHC in 2017--2018.
Tau leptons produced in heavy-flavor hadron decays and
\PW boson decays are exploited in the analysis.
The results from this analysis are combined with those of an earlier analysis using 2016 data,
which gives a combined total integrated luminosity of 131\fbinv. 
The observed (expected) upper limit on the branching fraction $\mathcal{B}(\tautrimu)$ is
$2.9\:(2.4) \times 10^{-8}$ at 90\% confidence level,
and $3.6\:(3.0) \times 10^{-8}$ at 95\% confidence level.
The result obtained in this search is the best from a hadron collider experiment, and comparable with the current most restrictive one from the Belle experiment.
As this limit is dominated by the statistical uncertainty, the additional data now being collected
will provide even more stringent tests of the standard model with this decay channel.

\begin{acknowledgments}
  We congratulate our colleagues in the CERN accelerator departments for the excellent performance of the LHC and thank the technical and administrative staffs at CERN and at other CMS institutes for their contributions to the success of the CMS effort. In addition, we gratefully acknowledge the computing centers and personnel of the Worldwide LHC Computing Grid and other centers for delivering so effectively the computing infrastructure essential to our analyses. Finally, we acknowledge the enduring support for the construction and operation of the LHC, the CMS detector, and the supporting computing infrastructure provided by the following funding agencies: SC (Armenia), BMBWF and FWF (Austria); FNRS and FWO (Belgium); CNPq, CAPES, FAPERJ, FAPERGS, and FAPESP (Brazil); MES and BNSF (Bulgaria); CERN; CAS, MoST, and NSFC (China); MINCIENCIAS (Colombia); MSES and CSF (Croatia); RIF (Cyprus); SENESCYT (Ecuador); MoER, ERC PUT and ERDF (Estonia); Academy of Finland, MEC, and HIP (Finland); CEA and CNRS/IN2P3 (France); SRNSF (Georgia); BMBF, DFG, and HGF (Germany); GSRI (Greece); NKFIH (Hungary); DAE and DST (India); IPM (Iran); SFI (Ireland); INFN (Italy); MSIP and NRF (Republic of Korea); MES (Latvia); LAS (Lithuania); MOE and UM (Malaysia); BUAP, CINVESTAV, CONACYT, LNS, SEP, and UASLP-FAI (Mexico); MOS (Montenegro); MBIE (New Zealand); PAEC (Pakistan); MES and NSC (Poland); FCT (Portugal); MESTD (Serbia); MCIN/AEI and PCTI (Spain); MOSTR (Sri Lanka); Swiss Funding Agencies (Switzerland); MST (Taipei); MHESI and NSTDA (Thailand); TUBITAK and TENMAK (Turkey); NASU (Ukraine); STFC (United Kingdom); DOE and NSF (USA).
  
  \hyphenation{Rachada-pisek} Individuals have received support from the Marie-Curie program and the European Research Council and Horizon 2020 Grant, contract Nos.\ 675440, 724704, 752730, 758316, 765710, 824093, and COST Action CA16108 (European Union); the Leventis Foundation; the Alfred P.\ Sloan Foundation; the Alexander von Humboldt Foundation; the Science Committee, project no. 22rl-037 (Armenia); the Belgian Federal Science Policy Office; the Fonds pour la Formation \`a la Recherche dans l'Industrie et dans l'Agriculture (FRIA-Belgium); the Agentschap voor Innovatie door Wetenschap en Technologie (IWT-Belgium); the F.R.S.-FNRS and FWO (Belgium) under the ``Excellence of Science -- EOS" -- be.h project n.\ 30820817; the Beijing Municipal Science \& Technology Commission, No. Z191100007219010 and Fundamental Research Funds for the Central Universities (China); the Ministry of Education, Youth and Sports (MEYS) of the Czech Republic; the Shota Rustaveli National Science Foundation, grant FR-22-985 (Georgia); the Deutsche Forschungsgemeinschaft (DFG), under Germany's Excellence Strategy -- EXC 2121 ``Quantum Universe" -- 390833306, and under project number 400140256 - GRK2497; the Hellenic Foundation for Research and Innovation (HFRI), Project Number 2288 (Greece); the Hungarian Academy of Sciences, the New National Excellence Program - \'UNKP, the NKFIH research grants K 124845, K 124850, K 128713, K 128786, K 129058, K 131991, K 133046, K 138136, K 143460, K 143477, 2020-2.2.1-ED-2021-00181, and TKP2021-NKTA-64 (Hungary); the Council of Science and Industrial Research, India; ICSC -- National Research Center for High Performance Computing, Big Data and Quantum Computing, funded by the EU NexGeneration program (Italy); the Latvian Council of Science; the Ministry of Education and Science, project no. 2022/WK/14, and the National Science Center, contracts Opus 2021/41/B/ST2/01369 and 2021/43/B/ST2/01552 (Poland); the Funda\c{c}\~ao para a Ci\^encia e a Tecnologia, grant CEECIND/01334/2018 (Portugal); the National Priorities Research Program by Qatar National Research Fund; MCIN/AEI/10.13039/501100011033, ERDF ``a way of making Europe", and the Programa Estatal de Fomento de la Investigaci{\'o}n Cient{\'i}fica y T{\'e}cnica de Excelencia Mar\'{\i}a de Maeztu, grant MDM-2017-0765 and Programa Severo Ochoa del Principado de Asturias (Spain); the Chulalongkorn Academic into Its 2nd Century Project Advancement Project, and the National Science, Research and Innovation Fund via the Program Management Unit for Human Resources \& Institutional Development, Research and Innovation, grant B37G660013 (Thailand); the Kavli Foundation; the Nvidia Corporation; the SuperMicro Corporation; the Welch Foundation, contract C-1845; and the Weston Havens Foundation (USA).  
\end{acknowledgments}

\bibliography{auto_generated}
\cleardoublepage \appendix\section{The CMS Collaboration \label{app:collab}}\begin{sloppypar}\hyphenpenalty=5000\widowpenalty=500\clubpenalty=5000
\cmsinstitute{Yerevan Physics Institute, Yerevan, Armenia}
{\tolerance=6000
A.~Hayrapetyan, A.~Tumasyan\cmsAuthorMark{1}\cmsorcid{0009-0000-0684-6742}
\par}
\cmsinstitute{Institut f\"{u}r Hochenergiephysik, Vienna, Austria}
{\tolerance=6000
W.~Adam\cmsorcid{0000-0001-9099-4341}, J.W.~Andrejkovic, T.~Bergauer\cmsorcid{0000-0002-5786-0293}, S.~Chatterjee\cmsorcid{0000-0003-2660-0349}, K.~Damanakis\cmsorcid{0000-0001-5389-2872}, M.~Dragicevic\cmsorcid{0000-0003-1967-6783}, P.S.~Hussain\cmsorcid{0000-0002-4825-5278}, M.~Jeitler\cmsAuthorMark{2}\cmsorcid{0000-0002-5141-9560}, N.~Krammer\cmsorcid{0000-0002-0548-0985}, A.~Li\cmsorcid{0000-0002-4547-116X}, D.~Liko\cmsorcid{0000-0002-3380-473X}, I.~Mikulec\cmsorcid{0000-0003-0385-2746}, J.~Schieck\cmsAuthorMark{2}\cmsorcid{0000-0002-1058-8093}, R.~Sch\"{o}fbeck\cmsorcid{0000-0002-2332-8784}, D.~Schwarz\cmsorcid{0000-0002-3821-7331}, M.~Sonawane\cmsorcid{0000-0003-0510-7010}, S.~Templ\cmsorcid{0000-0003-3137-5692}, W.~Waltenberger\cmsorcid{0000-0002-6215-7228}, C.-E.~Wulz\cmsAuthorMark{2}\cmsorcid{0000-0001-9226-5812}
% ...
\par}
\vskip\cmsinstskip
\dag:~Deceased\\
$^{1}$Also at Yerevan State University, Yerevan, Armenia\\
$^{2}$Also at TU Wien, Vienna, Austria\\
% ...
\end{sloppypar}
\end{document}